\documentclass[submission, Phys]{SciPost}

\usepackage[utf8]{inputenc} % input encodings (allow UTF-8 input)
\usepackage[T1]{fontenc} 	% selecting font encodings (use 8-bit T1 fonts)
\usepackage[english]{babel} % multilingual support (English language/hyphenation)

\usepackage[bitstream-charter]{mathdesign}
\usepackage{geometry} 		% flexible and complete interface to document dimensions
\usepackage{amsmath} 		% math package (American Mathematical Society)
\usepackage{mathtools} 		% math package (fixes various deficiencies of amsmath)
\usepackage{float} 			% floating objects such as figures and tables
\usepackage{graphicx} 		% enhanced support for graphics
\usepackage{tabularx} 		% tabulars with adjustable-width columns
\usepackage{booktabs} 		% professional-quality tables
\usepackage{color, xcolor} 	% foreground and background colour management
\usepackage{pdfpages} 		% inclusion of external multi-page PDF documents
\usepackage{extarrows} 		% extra arrows beyond those provided in amsmath
\usepackage{multirow} 		% create tabular cells spanning multiple rows
\usepackage{multicol} 		% intermix single and multiple columns
\usepackage{enumitem} 		% control layout of itemize, enumerate, description
\usepackage{xspace} 		% define commands that appear not to eat spaces
\usepackage{stackrel} 		% enhancement to the \stackrel command
\usepackage{tikz} 			% create PostScript and PDF graphics
\usepackage{braket} 		% Dirac bra-ket notation
\usepackage{bm} 			% access bold symbols in maths mode
\usepackage{tensor} 		% typeset tensors (tensor-style super- and subscripts)
\usepackage{slashed} 		% slash through characters (Feynman slash notation)
\usepackage{siunitx} 		% SI units package (typesetting values with units)
\usepackage{lastpage} 		% reference last page
\usepackage{cite} 			% improved citation handling
\usepackage[normalem]{ulem} % package for underlining
\usepackage{fontawesome} 	% access to web-related icons
\usepackage{tocloft} 		% control over the typography of the TOC, LOF, and LOT
\usepackage{titlesec} 		% interface to sectioning commands (various title styles)
\usepackage{doi} 			% create correct hyperlinks for DOI numbers
\usepackage{hyperref} 		% hypertext links (handel cross-referencing commands)
\usepackage[most]{tcolorbox} 					% coloured and framed text boxes
\usepackage[nameinlink, capitalize]{cleveref} 	% intelligent cross-referencing
\usepackage[nottoc, notlot, notlof]{tocbibind} 	% add references/index/contents to TOC
\usepackage[ruled, vlined]{algorithm2e} 		% floating algorithm environment
\usepackage{makecell}
\usepackage[makeroom]{cancel}
\usepackage{feynmf}

\makeatletter
\def\BState{\State\hskip-\ALG@thistlm}
\makeatother

\makeatletter
\@ifundefined{pdfoutput}{}{\DeclareGraphicsRule{*}{mps}{*}{}}
\makeatother

\makeatletter
\DeclareRobustCommand*{\bfseries}{%
   \not@math@alphabet\bfseries\mathbf
   \fontseries\bfdefault\selectfont
   \boldmath
}
\makeatother

\hypersetup{
	pdftitle={Profile Likelihoods on ML-Steroids},
	pdfauthor={Heimel, Plehn, Schmal},
	colorlinks=true, 			% false: boxed links, true: colored links
	linkcolor={red!50!black}, 	% color of internal links (sections, pages, etc.)
	citecolor={blue!50!black}, 	% color of citation links (links to bibliography)
	urlcolor={blue!80!black} 	% color of URL links (external links)
} 
\DeclareSymbolFont{usualmathcal}{OMS}{cmsy}{m}{n}
\DeclareSymbolFontAlphabet{\mathcal}{usualmathcal}

\SetArgSty{textnormal}
\SetKwComment{Comment}{{\small\#}~}{}
\SetCommentSty{mycommfont}

\setitemize{itemsep=0pt, parsep=0pt} 				% adjust itemize environment
\setenumerate{itemsep=0pt, parsep=0pt} 				% adjust enumerate environment
\setitemize{itemsep=2pt,topsep=2pt,parsep=0pt,partopsep=0pt,leftmargin=*}
\setenumerate{itemsep=0pt,topsep=2pt,parsep=0pt,partopsep=0pt,labelindent=3pt,leftmargin=*}
\usepackage{amsmath}
 
\usepackage{amsthm} 		% math package (typesetting theorems using AMS style)
\theoremstyle{definition}
 
\definecolor{red_cb}{HTML}{e41a1c}
\definecolor{blue_cb}{HTML}{377eb8}
\definecolor{green_cb}{HTML}{4daf4a}
\definecolor{purple_cb}{HTML}{984ea3}
\definecolor{orange_cb}{HTML}{ff7f00}

\definecolor{EmeraldGreen}{HTML}{1ea78d}
\definecolor{EnglishRed}{HTML}{b02427}
\hypersetup{colorlinks=true,urlcolor=EmeraldGreen,citecolor=EmeraldGreen,linkcolor=EnglishRed}

\newcommand{\ie}{\text{i.e.}\;}

\newcommand{\sfitter}{\textsc{SFitter}\xspace}
\newcommand{\ope}{\mathcal{O}}
\newcommand{\lag}{\mathcal{L}}
\newcommand{\Dfb}{\mbox{$\raisebox{2mm}{\boldmath ${}^\leftrightarrow$}\hspace{-4mm} D$}}
\newcommand{\Dfba}{\mbox{$\raisebox{2mm}{\boldmath ${}^\leftrightarrow$}\hspace{-4mm} D^a$}}

\newcommand{\mwith}{\text{with}}
\newcommand{\mand}{\text{and}}

\newcommand{\normal}{\mathcal{N}}

\newcommand{\XLangle}{\Bigl\langle}
\newcommand{\XRangle}{\Bigr\rangle}
\newcommand{\XXLangle}{\biggl\langle}
\newcommand{\XXRangle}{\biggr\rangle}

\newcommand{\qqquad}{\qquad\quad}

\newcommand{\pois}{\text{Poiss}}
\newcommand\one{\leavevmode\hbox{\small1\normalsize\kern-.33em1}}
\newcommand{\tr}{\operatorname{Tr}}			% trace
\newcommand{\loss}{\mathcal{L}} 	% loss value

\newcommand{\pytorch}{\textsc{PyTorch}\xspace}

\newcommand{\adam}{\textsc{Adam}\xspace}

\newcommand{\arXiv}[2][]{%
	\ifthenelse{\equal{#1}{}}%
	{\href{http://arxiv.org/abs/#2}{arXiv:#2}}%
	{\href{http://arxiv.org/abs/#2}{arXiv:#2~[#1]}}}

\def\slashchar#1{\setbox0=\hbox{$#1$}           % set a box for #1
   \dimen0=\wd0                                 % and get its size
   \setbox1=\hbox{/} \dimen1=\wd1               % get size of /
   \ifdim\dimen0>\dimen1                        % #1 is bigger
      \rlap{\hbox to \dimen0{\hfil/\hfil}}      % so center / in box
      #1                                        % and print #1
   \else                                        % / is bigger
      \rlap{\hbox to \dimen1{\hfil$#1$\hfil}}   % so center #1
      /                                         % and print /
   \fi}

\newcommand{\tikznode}[2]{%
\ifmmode%
\tikz[remember picture,baseline=(#1.base),inner sep=0pt] \node (#1) {$#2$};%
\else
\tikz[remember picture,baseline=(#1.base),inner sep=0pt] \node (#1) {#2};%
\fi}

\def\mathswitchr#1{\relax\ifmmode{\mathrm{#1}}\else$\mathrm{#1}$\xspace\fi}
\def\mathswitch#1{\relax\ifmmode#1\else$#1$\xspace\fi}

\graphicspath{{./figs/}}

\begin{document}

\begin{center}{\Large \textbf{
Profile Likelihoods on ML-Steroids
}}\end{center}

\begin{center}
Theo Heimel\textsuperscript{1,2},
Tilman Plehn\textsuperscript{1,3},
and
Nikita Schmal\textsuperscript{1}
\end{center}

\begin{center}
{\bf 1} Institut f\"ur Theoretische Physik, Universit\"at Heidelberg, Germany
\\
{\bf 2} CP3, Universit\'e catholique de Louvain, Louvain-la-Neuve, Belgium
\\
{\bf 3} Interdisciplinary Center for Scientific Computing (IWR), Universit\"at Heidelberg, Germany 
\end{center}

\begin{center}
\today
\end{center}

% For convenience during refereeing: line numbers
%\linenumbers

\section*{Abstract}
{\bf Profile likelihoods, for instance, describing global SMEFT
  analyses at the LHC are numerically expensive to construct and
  evaluate. Especially profiled likelihoods are notoriously unstable and noisy. We show how modern numerical tools, similar to neural
  importance sampling, lead to a huge numerical improvement and allow
  us to evaluate the complete SFitter SMEFT likelihood in five hours on a single GPU.}

\vspace{10pt}
\noindent\rule{\textwidth}{1pt}
\tableofcontents\thispagestyle{fancy}
\noindent\rule{\textwidth}{1pt}
\vspace{10pt}

\clearpage
%%%%%%%%%%%%%%%%%%%%%%%%%%%%%%%%%%%%%%%%%%%%%%%%%%%
\section{Introduction}
\label{sec:intro}

With the shift of the main physics paradigm of the LHC towards
data-driven and bottom-up precision analyses, global SMEFT analyses
allow us to answer the key question \textsl{Does the LHC data agree
  with the Standard Model altogether?} The reason is that SMEFT covers
all sectors of the Standard Model and allows us to combine huge
numbers of rate and kinematic measurements in a theoretically
meaningful way. These analyses define the Run2 legacy in the
Higgs-gauge
sector~\cite{Ellis:2018gqa,Biekotter:2018rhp,Kraml:2019sis,Almeida:2021asy},
the top
sector~\cite{Buckley:2015lku,Hartland:2019bjb,Brivio:2019ius,Maltoni:2019aot,Aoude:2022deh},
both sectors
combined~\cite{Ellis:2020unq,Ethier:2021bye,Elmer:2023wtr,Celada:2024mcf}.

The problem with global analyses, and the likely reason why none
of the LHC experiments have published a proper global analysis, is that
sampling the likelihood over the common space of Wilson coefficients
and nuisance parameters is extremely CPU-intensive, especially when we
want to avoid simplified Gaussian distributions. Removing nuisance
parameters and unwanted Wilson coefficients through profiling, rather
than the numerically easier marginalization, adds to the numerical misery. 

The first way to improve the likelihood evaluation is to use the structure of SMEFT as a perturbative
theory to simplify this sampling problem significantly. Secondly, 
modern machine learning and automatic differentiation allows
us to sample the likelihood in a parallelized and efficient
manner. Finally, neural importance
sampling~\cite{Gao:2020vdv,Danziger:2021eeg,Heimel:2022wyj,Deutschmann:2024lml},
for instance implemented in
MadNIS~\cite{Heimel:2022wyj,Heimel:2023ngj,Heimel:2024wph}, allows us
to sample the likelihood much more efficiently. 

In this paper we study
for the first time how a \sfitter global SMEFT analysis can be accelerated 
by combining these three directions.
%by including SMEFT features and at the same time adapting
%part of the MadNIS structures to likelihood sampling and the profiling
%of the fully exclusive likelihood.
In Sec.~\ref{sec:five} we review the construction of the \sfitter
likelihood and its perturbative structure and propose five steps to
numerical happiness. In Sec.~\ref{sec:smeft} we briefly summarize the construction of the \sfitter likelihood, such that we illustrate the five steps for a simplified 
toy model in Sec.~\ref{sec:res_toy} and then show that they also work
for the full SMEFT analysis in
Sec.~\ref{sec:res_lhc}. For the first time, this new method allows us to extract all Wilson coefficients for the Higgs-gauge and top sectors including the full correlated set of uncertainties in a numerically stable manner. In general, our numerical improvements allow us to 
compute SMEFT profile likelihoods in a few hours on a single GPU, rather than on
a CPU cluster over days and weeks.

%%%%%%%%%%%%%%%%%%%%%%%%%%%%%%%%%%%%%%%%%%%%%%%%%%%
\section{The five steps to happiness}
\label{sec:five}

The physics task behind our improved likelihood
sampling is a global SMEFT analysis, combining the top and Higgs-gauge
sectors including a comprehensive uncertainty treatment. With established methods, this analysis is possibly, but extremely
CPU-intensive. In Sec.~\ref{sec:five_like} we will describe the
structure of the fully exclusive likelihood and the challenges in
evaluating it fast. In Sec.~\ref{sec:five_sample} we will introduce
the five steps to sampling happiness.

%%%%%%%%%%%%%%%%%%%%%%%%%%%%%%%%%%%%%%%%%%%%%%%%%%%
\subsection{Constructing the likelihood}
\label{sec:five_like}

For decades, the exclusive likelihood for a single measurement as a
function of model parameters $c$ in \sfitter has been constructed
as~\cite{Lafaye:2004cn}
\begin{align}
  L_\text{excl}(\theta)
  = \pois(d|p(c, \theta, b)) \;
  \pois(b_{CR}|bk)\prod_i \,
  \mathcal{C}_{i}(\theta_{i}, \sigma_{i}) \; .
  \label{eqn:Excl_Likeli}
\end{align}
It includes the Poisson probability to observe $d$ events with $p$
events predicted, where the prediction is affected by uncertainties
encoded in nuisance parameters $\theta$. The backgrounds $b$ are
determined by an event count $b_{CR}$ in the control region and an
appropriate interpolation $k$ to the signal region. The constraints
$C_{i}$ describe the distributions of the nuisance parameters
$\theta_{i}$ with corresponding width $\sigma_{i}$.

%%%%%%%%%%%%%%%%%%%%%%%%%%%%%%%%%%%%%%%%%%%%%%%%%%%
\subsubsection*{Top sector}

For the top fit~\cite{Brivio:2019ius,Elmer:2023wtr} we use signal
strengths instead of rate measurements. Each measurement is then
modeled as a Gaussian rather than a Poisson distribution.  For
unfolded data, backgrounds do not need to be taken into account, so
the likelihood becomes a product of Gaussians,
\begin{align}
    L_\text{excl}(\theta) = \mathcal{N}(d|p(c, \theta))\prod_{i} \mathcal{C}_{i}(\theta_{i}, \sigma_{i}) \; .
\end{align}
The nuisance parameters should be removed through their constraints,
which depends on the type of underlying uncertainty. Systematic
uncertainties typically correspond to auxiliary measurements with
large amounts of data, so they are described by Gaussians. They
include, for instance, the luminosity or the lepton and photon
reconstruction. Theory uncertainties, for instance arising from
unknown higher orders or PDFs, are described by flat
likelihoods~\cite{Ghosh:2022lrf}.  We can remove the nuisance
parameters $\theta$ by profiling,
\begin{align}
  L_\text{prof}(\theta) = \max_\theta \ \mathcal{N}(d|p(c, \theta))
  \prod_i \, \mathcal{N}(\theta_{i},\sigma_{i}) \;
  \prod_j \, \mathcal{F}(\theta_{j},\sigma_{j}) \; .
\end{align}
The product of Gaussians becomes a single Gaussian with total width
$\sigma_\text{syst} = \sum_{i} \sigma_{syst, i}^{2}$, while the theory
uncertainties lead to a shift of the theory prediction by
$\sigma_\text{theo} = \sum_{i}\sigma_{theo,i}$ towards the data. By
profiling, theory uncertainties are added linearly, while systematics
are added quadratically. The final likelihood for signal strengths
becomes
\begin{align}
    \sqrt{ -2 \, \log L_\text{prof}(\theta)} = \begin{cases}
        (p + \sigma_{\text{theo}} - d)/\sigma_\text{syst} & d < p - \sigma_\text{theo}\\
        0 & d \in [p-\sigma_\text{theo}, p + \sigma_\text{theo}]\\
        (p -  \sigma_\text{theo} - d)/\sigma_\text{syst} & d > p + \sigma_\text{theo} \; .
    \end{cases}
\end{align}
%

%%%%%%%%%%%%%%%%%%%%%%%%%%%%%%%%%%%%%%%%%%%%%%%%%%%
\subsubsection*{Higgs-gauge sector}

The Higgs and di-boson data also includes measurements of signal plus
background rates~\cite{Butter:2016cvz,Biekotter:2018ohn}. Now, the
exclusive likelihood in Eq.\eqref{eqn:Excl_Likeli} cannot be
simplified, and we need to profile over a product of Poisson, Gaussian
and flat distributions.  If we are interested in the observed signal number $s =
d - b$, we can express the corresponding Poisson terms as a function
of $\tilde{s} = p - b$
\begin{align}
    \pois(d|p) = \pois(\tilde{s}|d,bk) = \frac{e^{-(\tilde{s}+bk)}(\tilde{s}+bk)^{d}}{d!} \; .
\end{align}
Furthermore, we define a generalized $\chi^{2}$, which vanishes when data and prediction match, 
\begin{align}
    \chi^{2} &= -2 \, \log 
    \frac{\pois(\tilde{s}|d,bk)\pois(b_{CR}|bk)}{\pois(\tilde{s}|p,bk)\pois(bk|bk)} \notag  \\
    &= -2 \left[ (d-p)\log(\tilde{s} + bk) + (b_{CR} - bk)\log(bk) + \log\left(\frac{p!}{d!}\frac{(bk)!}{b_{CR}!}\right) \right] \; .
\end{align}
Next, we profile over the expected background $b$. Performing this
maximization for each data point would be inefficient, we approximate
this contribution by splitting it into two parts
\begin{align}
    \log \, L_{\pois,d}(\Tilde{s}|d,b_{CR}) 
    &= d - (\tilde{s}_{\sigma} 
    + b_{CR}) \log (\tilde{s}_{\sigma} + b_{CR}) 
    + \log \frac{(\tilde{s}_{\sigma}+b_{CR})!}{d!} \notag \\
    \log \, L_{\pois,b}(\Tilde{s}|d,b_{CR}) 
    &= b_{CR} - (d - \tilde{s}_{\sigma}) \log (d - \tilde{s}_{\sigma}) 
    + \log \frac{(d - \tilde{s}_{\sigma})!}{b_{CR}!} \; .
\end{align}
We incorporate the effect of the flat nuisance parameters
by introducing the shifted signal $\tilde{s}_{\sigma} = \tilde{s} \pm
\sigma_\text{theo}$, where the sign is chosen such that the signal is
shifted towards data. The last missing piece are the Gaussian
systematics. These are analogous to the signal strengths and can be
computed as
\begin{align}
    -2 \, \log \, L_\text{Gauss}(\Tilde{s}|d,b_{CR}) 
    = \frac{(d - b_{CR} - \tilde{s}_{\sigma})^{2}}{\sum_\text{syst} (\sigma_{d,i} - \sigma_{b,i})^{2}} \, .
\end{align}
Finally, we combine all these contributions using the approximate
formula
\begin{align}
    \frac{1}{L_\text{full}} \approx \frac{1}{L_\text{Gauss}} + \frac{1}{L_{\text{Poiss},b}} + \frac{1}{L_{\text{Poiss},d}} \, ,
\end{align}
which is exact in the fully Gaussian case and has been shown to give
excellent results in previous \sfitter analyses~\cite{Lafaye:2009vr}.

%%%%%%%%%%%%%%%%%%%%%%%%%%%%%%%%%%%%%%%%%%%%%%%%%%%
\subsubsection*{Correlations}

A vital aspect of the \sfitter likelihood is that all systematic
uncertainties of the same type and experiment are fully correlated
between measurements. This is done through a correlation matrix with
off-diagonal entries
\begin{align}
    C_{ij} = \frac{\sum_{\text{syst}} \rho_{ij} \sigma_{i,\text{syst}}\sigma_{j,\text{syst}} }{\sigma_{i,\text{exp}}\sigma_{j,\text{exp}}} 
    \qquad  \text{with} \qquad 
    \sigma_{i,\text{exp}}^{2} = \sum_{\text{syst}}\sigma_{i,\text{syst}}^{2} + \sum_{\pois}\sigma_{i,\pois}^{2} \, ,
\end{align}
where the indices $i,j$ run over all measurements, and we choose
$\rho_{ij} = 0.99$ to ensure the invertibility of the correlation
matrix.
%The final likelihood is computed via
%%
%\begin{align}
%    \chi^{2} = \vec{\chi}_{i}^{T}C_{ij}^{-1}\vec{\chi}_{j} \, .
%\end{align}

%%%%%%%%%%%%%%%%%%%%%%%%%%%%%%%%%%%%%%%%%%%%%%%%%%%
\subsubsection*{Fast GPU-evaluation}

Computationally, the most costly part of computing likelihoods is
predicting the rate as a function of the Wilson coefficients. Using
the quadratic dependence of the rate on the Wilson coefficients we can
build suitable matrices to write the rates as a bilinear operation,
\begin{align}
    p^{(b)}_i = W_{ijk} C^{(b)}_j \tilde{C}^{(b)}_k + B_i
\end{align}
where $b$ is a batch index, $i$ runs over all observations, and $j$
and $k$ over the Wilson coefficients. $\tilde{C}$ is padded with a 1
to allow for linear dependencies. As this is a very simple tensor
operation, it can be accelerated considerably by moving it to a
GPU. Similar optimizations can be applied to other parts of the
likelihood, like the prediction of branching ratios in the Higgs
fit. By performing the computation in \pytorch, we also get access to
its gradients which allows us to use more efficient maximization
strategies.

%%%%%%%%%%%%%%%%%%%%%%%%%%%%%%%%%%%%%%%%%%%%%%%%%%%
\subsection{Improved likelihood sampling}
\label{sec:five_sample}

Neural importance sampling (NIS) offers a way to sample much more
efficiently from the \sfitter likelihood. However, likelihood sampling
defines a different set of challenges compared to MadNIS applied for
phase space
sampling~\cite{Heimel:2022wyj,Heimel:2023ngj,Heimel:2024wph}: the
likelihoods can be strongly correlated and multi-modal, and we do not
have access to physics-defined mappings and multi-channeling. Instead,
we account for these properties during the training of the normalizing
flow. Furthermore, sampling alone is not sufficient to get smooth
profile-likelihoods, so it has to be combined with a maximization
procedure. We divide our novel training and fitting method into five
steps:
\begin{enumerate}
    \item \textbf{Pre-scaling}: obtain an approximate sample from the
      likelihood to estimate component-wise means and
      standard-deviations;
    \item \textbf{Pre-training}: use that sample to train a
      normalizing flow to give the NIS training a better starting
      point;
    \item \textbf{Training}: run a NIS training, including annealed
      importance sampling and buffered training to improve the
      convergence;
    \item \textbf{Sampling}: use the trained normalizing flow to
      generate weighted samples. Make histograms and keep track of the
      maximal likelihood in each bin;
    \item \textbf{Maximizing}: use gradients to further improve the
      estimate of the profile likelihood computed during sampling.
\end{enumerate}
%

%%%%%%%%%%%%%%%%%%%%%%%%%%%%%%%%%%%%%%%%%%%%%%%%%%%
\subsubsection*{Pre-scaling}

To efficiently sample from a distribution, we shift and scale the
parameter space such that it is centered around zero with unit
standard deviation. This helps for MCMC as well as ML sampling. To
this end, we draw a batch of samples that approximates our target
distribution. As standard MCMC requires a burn-in and can only be
parallelized as independent Markov chains, we use annealed importance
sampling~\cite{neal2001annealed}.  It combines the advantages of
Markov chains and importance sampling by gradually transforming a
tractable base distribution $p_0(x)$ to the target $p_T(x)$ through
intermediate log-linear distributions
\begin{align}
    \log p_t(x) = (1-\beta_t) \log p_0(x) + \beta_t \log p_T(x)
    \qquad\mwith\qquad
    \beta_t = \frac{t}{T}
    \quad \text{and} \quad
    t = 1, \ldots, T    \; ,
\end{align}
We initialize the sampling
by drawing samples from a Gaussian base distribution,
\begin{align}
  x_0 \sim p_0(x_0) = \normal_{0,\sigma}(x_0)
  \qquad\mand\qquad
  w_0 = 1 \; .
\end{align}
Its standard deviation $\sigma$ should not be much narrower than the
target distribution, but does not require much tuning otherwise. Only in
cases where the width of the distribution differs by orders of magnitude
in different directions, it is necessary to coarsely initialize the scaling
by hand. Then the following steps are repeated for all $t$
\begin{enumerate}
\item transport the samples to the next $t$-distribution through
  re-weighting with
  \begin{align}
    \label{eq:ais_weights}
    w_t = \frac{p_t(x_{t-1})}{p_{t-1}(x_{t-1})} w_{t-1} \; ,
  \end{align}
\item evolve the samples according to the distribution $p_t$
      using one or more Metropolis-adjusted MCMC steps, $x_t =
      \text{MCMC}(x_{t-1})$ \; .
\end{enumerate}
At the end, we arrive at weighted samples $x_T$ with weights
\begin{align}
    w \equiv w_T = \prod_{t=1}^T \frac{p_t(x_{t-1})}{p_{t-1}(x_{t-1})} \; .
\end{align}
To get weights close to one, the number of $t$-steps has to be
sufficiently large. Because the samples are uncorrelated, the method
can be easily parallelized. For our MCMC steps we use a
%
%\begin{align}
%  x' = x + \sqrt{2\tau} \xi
%  \qquad \mwith \qquad
%  \xi \sim \normal_{0,1}(\xi) \; ,
%\end{align}
%
%corresponding to the
Gaussian proposal distribution with step size $\tau$,
\begin{align}
    q(x'|x) \propto \exp\left[ - \frac{(x' - x)^2}{4\tau} \right] \; .
\end{align}
We could use the available gradient information about our target
distribution for more sophisticated methods, like Langevin or
Hamiltonian Markov chains, but the additional cost of evaluating the
gradient is not justified by the improvement in sampling.  To ensure
detailed balance or unbiased sampling from the distribution $p$, we
use the Metropolis-Hastings algorithm and accept samples with the
probability
\begin{align}
    P_\text{accept}(x',x) &= \min \left[ 1, \:
    \frac{p(x') \: q(x|x')}{p(x) \: q(x'|x)} \right] \; .
\end{align}
The compromise between acceptance $a$ and parameter exploration
is determined by the step size $\tau$.  Since we do not use the samples
generated during pre-scaling later, we can make the step size adaptive
without having to worry about biasing our samples. After every MCMC
step and for a given target acceptance $a_\text{target}$, we update the
step size as
\begin{align}
    \tau \leftarrow \tau \times 2^{\min(r, 1)} \quad\mwith\quad r = \frac{a}{a_\text{target}} - 1 \;.
\end{align}
To improve the weights of the generated samples, we can use
re-sampling~\cite{del1997nonlinear} between steps 1 and 2, whenever
the effective sample size
\begin{align}
    N^\text{eff}_t = \frac{\left(\sum_i w_t^i\right)^2}{\sum_i (w_t^i)^2}
\end{align}
drops below a threshold. In that case we draw new samples from the
weighted samples $(w_t^i,x_t^i)$, using the normalized weights as
probabilities. After re-sampling, all samples are assigned the same
weight,
\begin{align}
    {w'}_t^i = \frac{1}{N} \sum_{i=1}^N w_t^i \; .
\end{align}
While this can initially lead to some degeneracy, the MCMC steps allow
samples to move away from their common starting point. This allows the
method to focus more on promising samples and leads to a much narrower
weight distribution.

At the end of the pre-scaling, we are left with weighted samples
$(w^i, x^i)$. We use these to calculate the component-wise means
$\mu_k$ and standard deviations $\sigma_k$, and from now on use the
standardized space coordinates $(x^i_k - \mu_k)/\sigma_k$.
%
%\begin{align}
%    x^i_k \leftarrow \frac{x^i_k - \mu_k}{\sigma_k} \; . 
%\end{align}

%%%%%%%%%%%%%%%%%%%%%%%%%%%%%%%%%%%%%%%%%%%%%%%%%%%
\subsubsection*{Flow pre-training}

In MadNIS we combine neural importance sampling with pre-defined
phase-space mappings, which incorporate our physics knowledge. To
sample over Wilson coefficients we do not have such mappings.

A standard NIS training, where we throw samples into the parameter
space and then optimize the network using these samples, will now fail
or converge very slowly.  Instead, we use a small set of samples from
our target distribution to pre-train the flow. For this purpose we can
use the samples from the pre-scaling. We perform a final re-sampling
on them and evolve them for a number of MCMC steps to get unweighted
training data. This way we avoid working with a too small pre-training
sample. The flow is then trained on these samples using a standard
log-likelihood loss,
\begin{align}
    \loss = -\log g_\theta(x) \;,
\end{align}
where $g_\theta(x)$ is the tractable probability distribution encoded
by the normalizing flow with trainable parameters $\theta$. Because we
typically perform the pre-training on a relatively small datasets,
like 10 batches of 1024 samples, we evolve the samples using further
MCMC steps after every batch.

While a short pre-training done this way is not sufficient to learn
the target distribution with high precision, it is sufficient to give
the main NIS training a sufficiently good starting point.

%%%%%%%%%%%%%%%%%%%%%%%%%%%%%%%%%%%%%%%%%%%%%%%%%%%
\subsubsection*{Flow training}

Annealed importance sampling can also be used to make the training of
flow networks for neural importance sampling more efficient and avoid
common problems like mode collapse or inefficient training for
distributions with narrow features~\cite{midgley2022flow}. Consider a
variance loss
\begin{align}
    \label{eq:variance_loss}
    \loss = \left\langle \frac{p(x)^2}{g_\theta(x) q(x)} \right\rangle_{x \sim q(x)}
\end{align}
with a proposal distribution $q(x)$. In a regular NIS training we
choose $q(x) = g_\theta(x)$, such that we can use our flow to generate
the training samples online. However, this proposal does
not minimize the variance of the loss. \ie the variance of
the variance. The optimal proposal distribution is
\begin{align}
    \label{eq:opt_proposal}
    q(x) = f_\theta(x) \equiv \frac{p(x)^2}{g_\theta(x)} \; ,
\end{align}
such that the expectation value in Eq.\eqref{eq:variance_loss} becomes
trivial. We can get weighted samples from this distribution using
annealed importance sampling, using the flow
$g_\theta(x)$ as the base distribution and $f_\theta(x)$ as the target
distribution. After setting $q(x) = f_\theta(x)$ in
Eq.\eqref{eq:variance_loss}, we can rewrite the gradients as
\begin{align}
    \nabla_\theta \loss
    = \XXLangle \frac{\nabla_\theta f_\theta(x)}{f_\theta(x)} \XXRangle_{x \sim q(x)}
    = \XLangle \nabla_\theta \log f_\theta(x) \XRangle_{x \sim q(x)}
    = - \XLangle \nabla_\theta \log g_\theta(x) \XRangle_{x \sim q(x)} \; .
\end{align}
After annealed importance sampling and in terms of its weights $w$, we
can write the loss as
\begin{align}
    \loss_\text{online} = - \XLangle w \log g_\theta(x) \XRangle_{x \sim g_\theta(x)} + \text{const} \; .
\end{align}
To
improve the stability of this loss evaluation, we apply a batch-wise
normalization of the weights and limit the effect of very large
weights using a modified weight function, which is approximately linear with unit slope
for small weights and logarithmic for $\gtrsim 10$, 
\begin{align}
    w' = \alpha \log\left(\frac{w}{\alpha}+1\right) \quad\mwith\quad \alpha=30 \;.
\end{align}

As in MadNIS, we speed up the training using a buffered sampling
step. However, instead of training the network on weighted samples
taken from the buffer with uniform probability, we instead follow the
method proposed in Ref.~\cite{midgley2022flow}. We draw samples from
the buffer with a probability proportional to their importance
sampling weight, such that the weight update is performed on
approximately unweighted events. We only have to account for the
change in the network parameters at the time of sampling $\theta'$ and
at the time of optimization $\theta$. The loss function of a buffered
training step then reads
\begin{align}
  \loss_\text{buffered}
  = \XXLangle \frac{q_\theta(x)}{q_{\theta'}(x)} \log q_\theta(x) \XXRangle \; .
%    \quad\mwith\quad w_\text{corr} =  \;,
\end{align}
After each buffered sampling step, we update the buffered weights
using the $q$-ratio to reduce the bias of the
buffered samples compared to the target distribution.

%%%%%%%%%%%%%%%%%%%%%%%%%%%%%%%%%%%%%%%%%%%%%%%%%%%
\subsubsection*{Sampling}

After the network is trained, we can use it to draw weighted samples
from our target distribution. We are interested in 1D and 2D marginal
distributions which we generate as histograms. Furthermore, we are
interested in the 1D and 2D profile likelihoods, \ie the maximum of
the likelihood when one or two model parameters are fixed.

We fix the binning of the histograms for marginalization and profiling
after the first batch of generated samples. We then generate 1D
histograms for all parameters and 2D histograms for all parameter
pairs.  For the profile likelihood, we select the sample with the
highest likelihood in each bin for each batch of samples. We then
change the one or two fixed parameters such that the point is in the
center of the bin and re-evaluate the likelihood for the updated
parameter point. We then choose a fixed number of points with the
highest likelihoods for each bin from all batches.

%%%%%%%%%%%%%%%%%%%%%%%%%%%%%%%%%%%%%%%%%%%%%%%%%%%
\subsubsection*{Maximizing}

The sampling provides us with points close to the maximum likelihood,
with one or two parameters fixed. To further improve the profile
likelihood and to reduce the effects from limited statistics, we use
these points as the starting point for a maximization based on
gradient-ascent through automatic differentiation in \pytorch.

In simple cases it is sufficient to use fast gradient-ascent based
optimizers like \adam. For more complex likelihoods we use the L-BFGS optimizer, a second-order optimization method, as it allows for more
precise estimation of the maximum.

%%%%%%%%%%%%%%%%%%%%%%%%%%%%%%%%%%%%%%%%%%%%%%%%%%%
\section{SMEFT}
\label{sec:smeft}

The SMEFT~\cite{Buchmuller:1985jz,Leung:1984ni,Degrande:2012wf,Brivio:2017vri} provides a perfect framework for model agnostic searches for new physics arising from particles too heavy to be produced directly. We parametrize these new physics contributions via Wilson coefficients $C_{k}$ and higher-dimensional operators $\ope_k$. They are included in the SMEFT Lagrangian by expanding it in inverse powers of the new physics scale $\Lambda$, respecting the symmetries and field content of the SM 
\begin{align}
\lag_\text{SMEFT} = \lag_\text{SM} + \sum_k \frac{C_k}{\Lambda^2} \; \ope_k \;\;.
\end{align}
In our analysis we ignore lepton-number violating operators, which removes any dimension-5 operators, and CP-violating operators. To search for those we prefer dedicated, optimal analyses~\cite{Brehmer:2017lrt,Bahl:2021dnc,Barman:2021yfh}. %Without further assumptions there remain 2499 operators, far exceeding the number of available measurements. 
Moreover, we truncate our SMEFT expansion at dimension six, motivated by the assumption that the suppression in terms of $\Lambda$ translates from the Lagrangian to the physical observables by integrating out heavy particle contributions. For limitations induced by this truncations we refer for instance to Refs.~\cite{Dawson:2022cmu,Dawson:2024ozw}.

The number of operators can be further reduced by restricting the analysis to specific classes of processes. For our analysis we include top quark observables, as well as Higgs, di-boson and electroweak precision observables (EWPOs). SMEFT operator contributions are computed up to quadratic order for top, Higgs and di-boson observables, while only linear terms are included for EWPOs. Additional assumptions, such as those regarding flavor structure, depend on the type of data considered.

%%%%%%%%%%%%%%%%%%%%%%%%%%%%%%%%%%%%%%%%%%%%%%%%%%%
\subsection{Top sector}

%---------------------------------------
\begin{table}[t]
    \centering
    \begin{small} \begin{tabularx}{\textwidth}{lXlX}
    \toprule
        Operator & Definition & Operator & Definition \\
        \midrule
        $\ope_{Qq}^{1,8}$ & $(\bar{Q}\gamma_\mu T^A Q) \; (\bar{q}_i\gamma^\mu T^A q_i)$ &
        $\ope_{tu}^8$ & $(\bar{t}\gamma_\mu T^A t) \; (\bar{u}_i\gamma^\mu T^A u_i)$ \\
        $\ope_{Qq}^{1,1}$ & $(\bar{Q}\gamma_\mu Q) \; (\bar{q}_i\gamma^\mu q_i)$ &
        $\ope_{tu}^1$ & $(\bar{t}\gamma_\mu t) \; (\bar{u}_i\gamma^\mu u_i)$ \\
        $\ope_{Qq}^{3,8}$ & $(\bar{Q}\gamma_\mu T^A\tau^I Q) \; (\bar{q}_i\gamma^\mu T^A \tau^I q_i)$ &
        $\ope_{td}^8$ & $(\bar{t}\gamma^\mu T^A t) \; (\bar{d}_i\gamma_\mu T^A d_i)$ \\
        $\ope_{Qq}^{3,1}$ & $(\bar{Q}\gamma_\mu\tau^I Q) \; (\bar{q}_i\gamma^\mu\tau^I q_i)$ &
        $\ope_{td}^1$ & $(\bar{t}\gamma^\mu t) \; (\bar{d}_i\gamma_\mu d_i)$ \\
        \midrule
        $\ope_{Qu}^8$ & $(\bar{Q}\gamma^\mu T^A Q) \; (\bar{u}_i\gamma_\mu T^A u_i)$ &
        $\ope_{Qd}^1$ & $(\bar{Q}\gamma^\mu Q) \; (\bar{d}_i\gamma_\mu d_i)$ \\
        $\ope_{Qu}^1$ & $(\bar{Q}\gamma^\mu Q) \; (\bar{u}_i\gamma_\mu u_i)$ &
        $\ope_{tq}^8$ & $(\bar{q}_i\gamma^\mu T^A q_i) \; (\bar{t}\gamma_\mu T^A t)$ \\
        $\ope_{Qd}^8$ & $(\bar{Q}\gamma^\mu T^A Q) \; (\bar{d}_i\gamma_\mu T^A d_i)$ &
        $\ope_{tq}^1$ & $(\bar{q}_i\gamma^\mu q_i) \; (\bar{t}\gamma_\mu t)$ \\
        \midrule
        $\ope_{\phi Q}^{1}$ & $(\phi^\dagger\,i \stackrel{\longleftrightarrow}{D_\mu} \phi) \; (\bar{Q}\gamma^{\mu}Q)$ &
        $^\ddagger \ope_{tB}$ & $(\bar{Q}\sigma^{\mu\nu} t)\,\widetilde{\phi}\,B_{\mu\nu}$ \\
        $\ope_{\phi Q}^{3}$ & $(\phi^\dagger\,i \stackrel{\longleftrightarrow}{D_\mu^I} \phi) \; (\bar{Q}\gamma^{\mu}\tau^I Q)$ &
        $^\ddagger \ope_{tW}$ & $(\bar{Q}\sigma^{\mu\nu} t)\,\tau^I\widetilde{\phi}\,W_{\mu\nu}^I$ \\
        $\ope_{\phi t}$ & $(\phi^\dagger\,i \stackrel{\longleftrightarrow}{D_\mu} \phi) \; (\bar{t}\gamma^{\mu}t)$ &
        $^\ddagger \ope_{bW}$ & $(\bar{Q}\sigma^{\mu\nu} b)\,\tau^I\phi \,W_{\mu\nu}^I$ \\
        $^\ddagger \ope_{\phi tb}$ & $(\widetilde{\phi}^\dagger iD_\mu \phi) \; (\bar{t}\gamma^{\mu}b)$ &
        $^\ddagger \ope_{tG}$ & $(\bar{Q}\sigma^{\mu\nu} T^A t)\,\widetilde{\phi}\,G_{\mu\nu}^A$ \\
        \bottomrule
    \end{tabularx} \end{small}
    \caption{List of the 22 independent operators contributing to our top observables. They are related to the Warsaw basis in the Appendix of Ref.~\cite{Brivio:2019ius}.}
    \label{tab:top_operators}
\end{table}
%---------------------------------------

The operator basis for the top sector analysis follows Ref.~\cite{Brivio:2019ius}, with non-hermitian operators denoted as $^\ddagger \ope$. We impose a $U(2)$ flavor symmetry on the first and second quark generations, as most top observables are blind to the flavor of light quarks,
\begin{alignat}{5}
q_i & = (u^i_L,d^i_L) & \qqquad 
u_i & = u^i_R,\,d_i=d^i_R 
\quad \text{for} \quad i=1,2 \notag  \\
Q & = (t_L,b_L) & \qqquad 
t &= t_R,\,b=b_R \; .
\label{eq:symm}
\end{alignat}
All quark masses except for the top mass are taken to be zero, which leaves 22 independent operators listed in Tab~\ref{tab:top_operators}.

The top-sector operators are divided into three distinct categories. The top row describes four-fermion currents with $RR$ and $LL$ helicity structure, while the center row operators include a $RL$ and $LR$ helicity flips. These are mainly constrained by observables involving top pairs, such as $t\bar{t}$ and associated $t\bar{t}W, t\bar{t}Z$ production. The final set of operators couple heavy quarks to gauge bosons. For convenience, we make use of additional relations arising from gauge invariance
\begin{align}
C_{\phi Q}^- = C_{\phi Q}^1 - C_{\phi Q}^3 
\qquad \text{and} \qquad
C_{tZ} = c_w C_{tW} - s_w C_{tB}. 
\end{align}
With these definitions we choose $C_{\phi Q}^{-},C_{\phi Q}^{3},C_{tW}$ and $C_{tZ}$ as our degrees of freedom. To fully constrain them, we include single top production, associated $tZ$ and $tW$ production, as well as top decay observables. A detailed description of the observables included in our dataset and the impact of the different Wilson coefficients is given in Ref.~\cite{Elmer:2023wtr}. As part of this study we also showed that we can use published likelihoods by ATLAS and CMS, including detailed uncertainty information, in our \sfitter analysis.

%%%%%%%%%%%%%%%%%%%%%%%%%%%%%%%%%%%%%%%%%%%%%%%%%%%
\subsection{Higgs-gauge sector}

%---------------------------------------
\begin{table}[ht]
    \centering
    \begin{small} \begin{tabularx}{\textwidth}{lXlX}
    \toprule
         Operator & Definition & Operator & Definition \\
        \toprule
        $\ope_{GG}$ & $\phi^\dagger \phi \; G_{\mu\nu}^a G^{a\mu\nu}$ &
        $\ope_{WW}$ & $\phi^{\dagger} \; \hat{W}_{\mu \nu} \hat{W}^{\mu \nu} \; \phi$ \\
        $\ope_{BB}$ & $\phi^{\dagger} \; \hat{B}_{\mu \nu} \hat{B}^{\mu \nu} \; \phi$ &
        $\ope_W$ & $(D_{\mu} \phi)^{\dagger}  \hat{W}^{\mu \nu}  (D_{\nu} \phi)$ \\
        $\ope_B$ & $(D_{\mu} \phi)^{\dagger}  \hat{B}^{\mu \nu}  (D_{\nu} \phi)$ &
        $\ope_{BW}$ & $\phi^\dagger \; \hat{B}_{\mu\nu} \hat{W}^{\mu\nu} \; \phi$ \\
        $\ope_{\phi 1}$ & $(D_\mu \phi)^\dagger \; \phi \phi^\dagger \; (D^\mu \phi)$ &
        $\ope_{\phi 2}$ & $\frac{1}{2} \partial^\mu ( \phi^\dagger \phi ) \partial_\mu ( \phi^\dagger \phi )$ \\
        $\ope_{3W}$ & $\tr \left( \hat{W}_{\mu \nu} \hat{W}^{\nu \rho} \hat{W}_\rho^\mu \right)$ & & \\
        \midrule
        $\ope_{\phi u}^{(1)}$ & $\phi^\dagger (i\,{\Dfb}_{\mu} \phi) (\bar u_{R}\gamma^\mu u_{R})$ &
        $\ope_{\phi Q}^{(1)}$ & $\phi^\dagger (i\,{\Dfb}_{\mu} \phi) (\bar Q\gamma^\mu Q)$ \\
        $\ope_{\phi d}^{(1)}$ & $\phi^\dagger (i\,{\Dfb}_{\mu} \phi) (\bar d_{R}\gamma^\mu d_{R})$ &
        $\ope_{\phi Q}^{(3)}$ & $\phi^\dagger (i\,{\Dfba}_{\!\!\mu} \phi) \left(\bar Q\gamma^\mu \frac{\sigma_a}{2} Q\right)$ \\
        $\ope_{\phi e}^{(1)}$ & $\phi^\dagger (i\,{\Dfb}_{\mu} \phi) (\bar e_{R}\gamma^\mu e_{R})$ & & \\
        \midrule
        $\ope_{e\phi,22}$ & $\phi^\dagger\phi \; \bar L_2 \phi e_{R,2}$ &
        $\ope_{e\phi,33}$ & $\phi^\dagger\phi \; \bar L_3 \phi e_{R,3}$ \\
        $\ope_{u\phi,33}$ & $\phi^\dagger\phi \; \bar Q_3 \tilde \phi u_{R,3}$ &
        $\ope_{d\phi,33}$ & $\phi^\dagger\phi \; \bar Q_3 \phi d_{R,3}$ \\
        \midrule
        $\ope_{4L}$ & $(\bar{L}_1 \gamma_\mu L_2) \; (\bar{L}_2 \gamma^\mu L_1)$ & & \\
        \bottomrule
    \end{tabularx} \end{small}
    \caption{List of the 19 independent operators contributing to our Higgs, di-boson and electroweak observables.}
    \label{tab:higgs_operators}
\end{table}
%---------------------------------------

For the analysis of Higgs, di-boson and EWPOs data we adopt the HISZ operator basis based on physics arguments from Run1 and Run2~\cite{Butter:2016cvz,Biekotter:2018ohn,Brivio:2022hrb}. All  operators are listed in Tab~\ref{tab:higgs_operators}. They are, again, split into several categories. The top row lists all operators affecting the Higgs interactions with gauge bosons, all of which can be constrained using Higgs processes, except for $\ope_{3W}$, which is constrained by di-boson production and electroweak precision measurements. 
The next two rows list single-current operators affecting both gauge and Higgs-gauge couplings. These are split into two parts, where $\ope_{\phi u}^{1},\ope_{\phi d}^{1},\ope_{\phi e}^{1},\ope_{\phi q}^{1},\ope_{\phi q}^{3}$ are flavor universal, while we allow for minimal flavor violation coming from $\ope_{e\phi,22},\ope_{e\phi,33},\ope_{u\phi,33},\ope_{d\phi,33}$.
Finally, we include the four-lepton operator $\ope_{4L}$ which induces a shift in the Fermi constant.
As in the top analysis we use an orthogonal combination as our actual degrees of freedom
\begin{align}
    \ope_\pm = \frac{\ope_{WW} \pm \ope_{BB}}{2} \qquad \Rightarrow \qquad f_{\pm} = f_{WW} \pm f_{BB} \, .
\end{align}
This way only $\ope_{+}$ contributes to the $H\gamma\gamma$ interaction.

%%%%%%%%%%%%%%%%%%%%%%%%%%%%%%%%%%%%%%%%%%%%%%%%%%%
\subsection{Combined analysis}

It is not possible to naively combine our two top and Higgs-gauge  operator bases into a combined analysis, because they are based on different flavor assumptions. First, we need to convert the HISZ-operators in Tab.~\ref{tab:higgs_operators} into the Warsaw basis using the expressions derived in the Appendix of Ref.~\cite{Brivio:2021alv}. In the Warsaw basis, illustrated in Tab.~\ref{tab:higgs_operators_warsaw}, we can align the flavor structures of the two sectors. After that, the combined analysis including comprehensive and fully correlated uncertainties is just a numerical problem. 

On the physics side, the combination of the two sectors defines a few distinct bridges between the largely de-correlated sets of operators and observables. On the observable side, the obvious bridges are associated top (pair) production with Higgs and gauge bosons. For \sfitter, the historic split is to include $t \bar{t}H$ production in the Higgs-gauge sector, because it cannot be separated from gluon-fusion Higgs production. On the other hand, for instance $t\bar{t}Z$ production is part of the top sector analysis, leading to an artificial split motivating a combination of the two sectors. In addition to the top Yukawa operator, the main bridge between the two sectors is $\ope_{tG}$~\cite{Ellis:2020unq}, originally part of the \sfitter top sector analysis.

While the combination of the top and Higgs-gauge sectors to a proper global SMEFT analysis is a clear physics goal, it leads to technical complications. The requirements on the treatment of the likelihoods in the two sectors are very different. For the Higgs-gauge sector kinematic distributions, for instance in di-boson and $VH$ production, have shown to be extremely powerful and largely statistics-limited. Theory and systematic uncertainties and their correlations are less important in kinematic tails, but under-fluctuations in low-statistics regimes have to be accommodated, leading to significant differences between profiled and marginalized single-operator limits~\cite{Brivio:2022hrb}. In contrast, the top sector analysis is driven by (associated) top pair production, a QCD process with large rates. Even kinematic tails are heavily populated, and the correlated theory uncertainties are key to an otherwise mostly Gaussian analysis. Again, the combination of a flat theory uncertainty with a Gaussian nuisance parameter leads to sizeable differences between profiled and marginalized results. All of these differences pose technical requirements on the combined analysis, where the subtleties of both sectors have to be treated correctly.

%---------------------------------------
\begin{table}[t]
\centering
\begin{small} \begin{tabularx}{\textwidth}{lXlX}
\toprule
Operator & Definition & Operator & Definition\\
    \midrule
    $\ope_{\phi G}$ & $\phi^{\dagger} \phi G_{\mu \nu}^{A} G^{A \mu \nu}$ & 
    $\ope_{W}$ & $\varepsilon^{IJK} W_{\mu}^{I \nu} W_{\nu}^{J \rho} W_{\rho}^{K \mu}$ \\
    $\ope_{\phi B}$ & $\phi^{\dagger} \phi B_{\mu \nu} B^{\mu \nu}$ & 
    $\ope_{\phi W}$ & $\phi^{\dagger} \phi W_{\mu \nu}^{I} W^{I \mu \nu}$ \\
    $\ope_{\phi WB}$ & $\phi^{\dagger} \tau^{I} \phi W_{\mu \nu}^{I} B^{\mu \nu}$ & & \\
    \midrule
    $\ope_{\phi \Box}$ & $(\phi^{\dagger} \phi ) \Box (\phi^{\dagger} \phi )$ & 
    $\ope_{\phi D}$ & $(\phi^{\dagger} D^{\mu} \phi ) ^{*} (\phi^{\dagger} D^{\mu} \phi )$ \\
    $\ope_{\phi e}$ & $(\phi^{\dagger} i \overleftrightarrow{D_{\mu}} \phi) (\bar{e}_i \gamma^{\mu} e_i)$ & 
    $\ope_{\phi b}$ & $(\phi^{\dagger} i \overleftrightarrow{D_{\mu}} \phi) (\bar{b}_{i} \tau^{I} \gamma^{\mu} b_{i})$ \\
    $\ope_{\phi d}$ & $\sum_{i=1}^{2}(\phi^{\dagger} i \overleftrightarrow{D_{\mu}} \phi) (\bar{d}_i \gamma^{\mu} d_i)$ & 
    $\ope_{\phi u}$ & $\sum_{i=1}^{2}(\phi^{\dagger} i \overleftrightarrow{D_{\mu}} \phi) (\bar{u}_i \gamma^{\mu} u_i)$ \\
    $\ope_{\phi q}^{(1)}$ & $\sum_{i=1}^{2}(\phi^{\dagger} i \overleftrightarrow{D_{\mu}} \phi) (\bar{q}_i \gamma^{\mu} q_i)$ & 
    $\ope_{\phi q}^{(3)}$ & $\sum_{i=1}^{2}(\phi^{\dagger} i \overleftrightarrow{D_{\mu}} \phi) (\bar{q}_{i} \tau^{I} \gamma^{\mu} q_{i})$ \\
    $\ope_{\phi l}^{(1)}$ & $(\phi^{\dagger} i \overleftrightarrow{D_{\mu}} \phi) (\bar{l} \gamma^{\mu} l) $ & 
    $\ope_{\phi l}^{(3)}$ & $(\phi^{\dagger} i \overleftrightarrow{D}^{I}_{\mu} \phi) (\bar{l} \tau^{I} \gamma^{\mu} l)$ \\
    \midrule
    $\ope_{d \phi, 33}$ & $(\phi^{\dagger} \phi)(\bar{Q}_{3} b \phi)$ & 
    $\ope_{u \phi, 33}$ & $(\phi^{\dagger} \phi)(\bar{Q}_{3} t \phi)$ \\
    $\ope_{e \phi, 22}$ & $(\phi^{\dagger} \phi)(\bar{l}_{2} \mu \phi)$ & 
    $\ope_{e \phi, 33}$ & $(\phi^{\dagger} \phi)(\bar{l}_{3} \tau \phi)$ \\
    \midrule
    $\ope_{ll}$ & $(\bar{l} \gamma_{\mu} l)(\bar{l} \gamma^{\mu} l)$ & & \\
    \bottomrule
\end{tabularx} \end{small}
\caption{Operators in the Warsaw basis for Higgs, di-boson and electroweak sectors. These 21 degrees of freedom and those from the Top sector are included in the combined global analysis.}
\label{tab:higgs_operators_warsaw}
\end{table}
%---------------------------------------

%%%%%%%%%%%%%%%%%%%%%%%%%%%%%%%%%%%%%%%%%%%%%%%%%%%
\section{Results}
\label{sec:res}

The focus of this paper is not a new SMEFT analysis, but a new combination of physics assumptions and ML-methods to evaluate the fully exclusive likelihood and extract marginalized and profiled likelihoods. We start by illustrating this technique for a toy example in Sec.~\ref{sec:res_toy} and then show the improved global \sfitter analysis in Sec.~\ref{sec:res_lhc}.

%%%%%%%%%%%%%%%%%%%%%%%%%%%%%%%%%%%%%%%%%%%%%%%%%%%
\subsection{Toy example}
\label{sec:res_toy}

%----------------------------------------------------------
\begin{figure}[b!]
    \includegraphics[width = 0.495\textwidth]{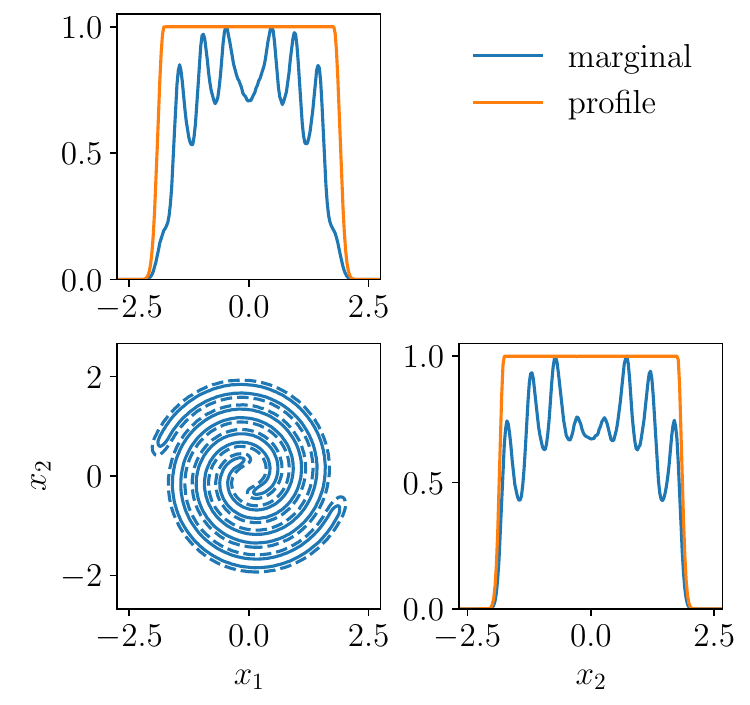}
    \includegraphics[width = 0.495\textwidth]{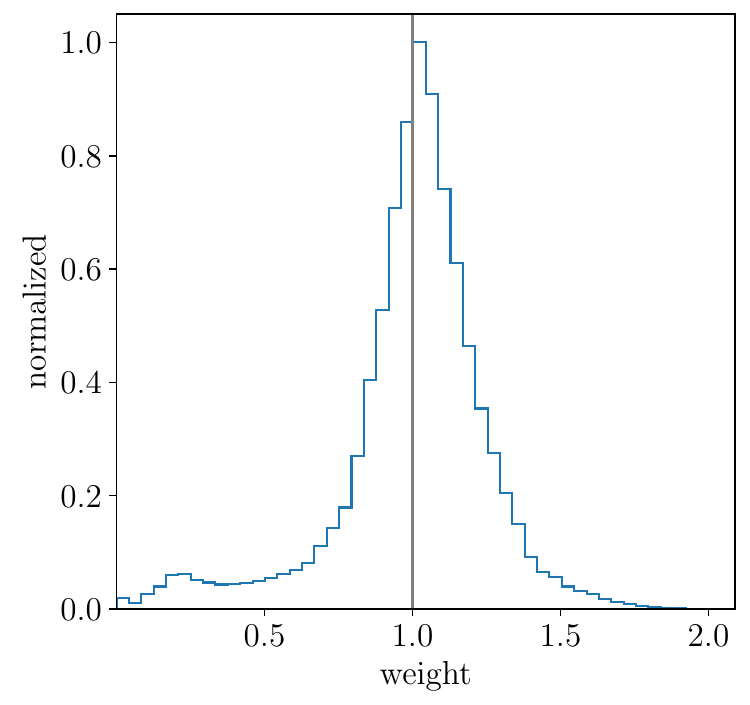}\\
    \includegraphics[width = 0.495\textwidth]{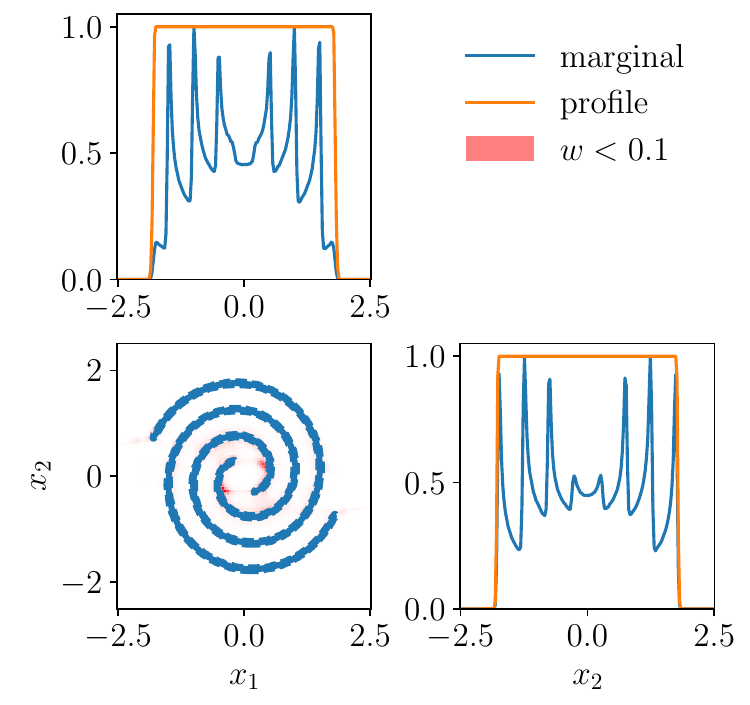}
    \includegraphics[width = 0.495\textwidth]{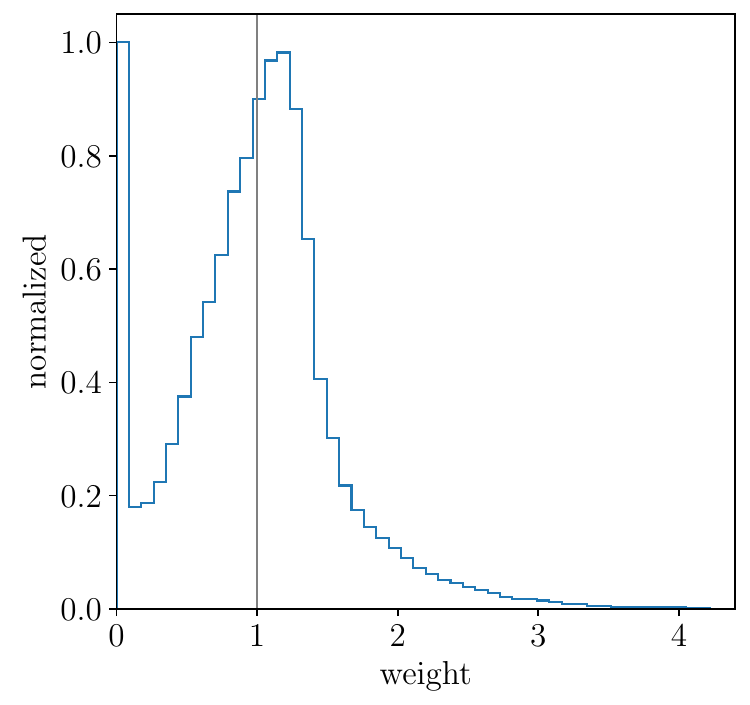}
    \caption{Profiled and marginalized likelihoods, 68\% and 95\%
      confidence regions, and weight distributions for the spiral toy
      distribution with widths $\sigma=0.007$ (upper) and
      $\sigma=0.0005$ (lower).}
    \label{fig:spiral}
\end{figure}
%----------------------------------------------------------

As a first toy model we define two two-dimensional concentric spirals as
\begin{align}
  \begin{pmatrix} r \\ \phi \end{pmatrix}
  = \pm \begin{pmatrix} t \\ 2 \pi t \end{pmatrix}
  \quad \text{with} \quad
  t \in [0.36, 1.93] \; .
\end{align}
We turn this spiral into a probability distribution by imposing a
uniform distribution along the length of the spiral, and smearing with
a Gaussian distribution with a constant width $\sigma$. We then
perform training, marginalization and profiling for this distribution
with the hyperparameters given in the Appendix. The results for two
Gaussian widths are shown in Fig.~\ref{fig:spiral}.
%We only show the
%profiled results in the one-dimensional histograms, as the profiled
%and marginalized distributions are identical in the two-dimensional
%case.
Sampling and profiling work well in both cases, and
the ML-sampler has no problems resolving the complex structure of the
spiral. The gradients in the maximization pass
give us a completely smooth profiled likelihood.

For the narrow width, we see a secondary peak in the weight
distribution for small weights. This is caused by the network
interpolating into the low-probability regions of the distribution in
places where it is not able to learn the distribution perfectly. We
confirm this by plotting the distribution of samples with $w < 0.1$ in
the bottom left panel of Fig.~\ref{fig:spiral}.

%----------------------------------------------------------
\begin{figure}[t]
    \includegraphics[width = 0.495\textwidth]{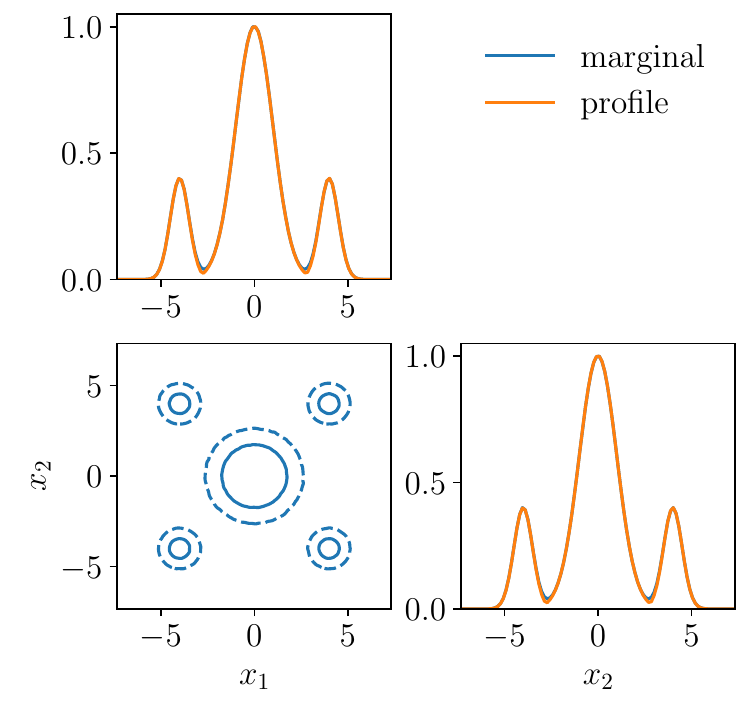}
    \includegraphics[width = 0.495\textwidth, page=1]{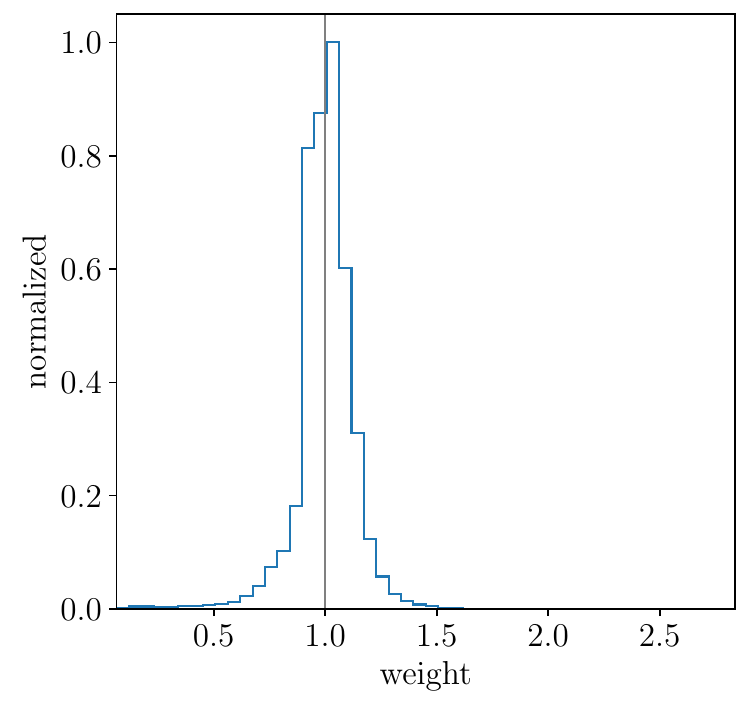}\\
    \includegraphics[width = 0.495\textwidth]{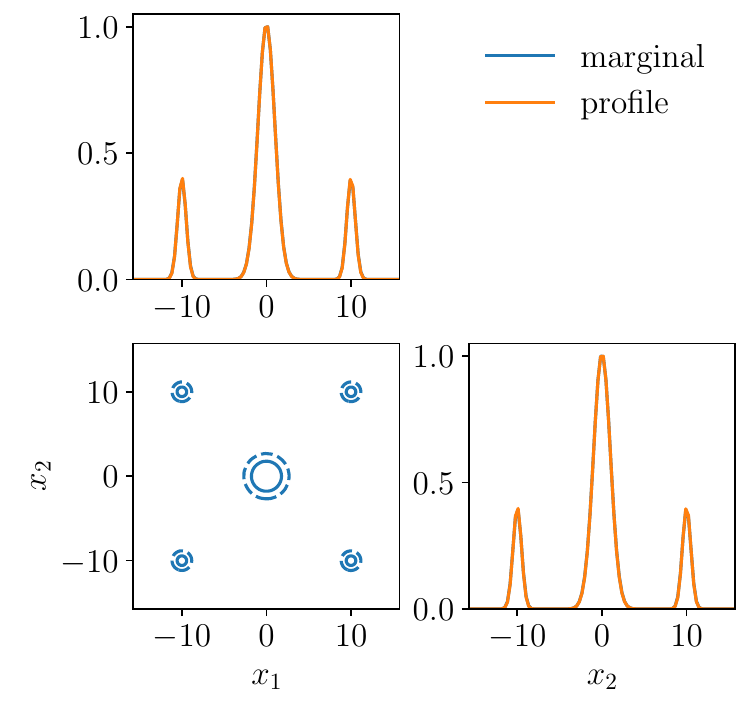}
    \includegraphics[width = 0.495\textwidth, page=1]{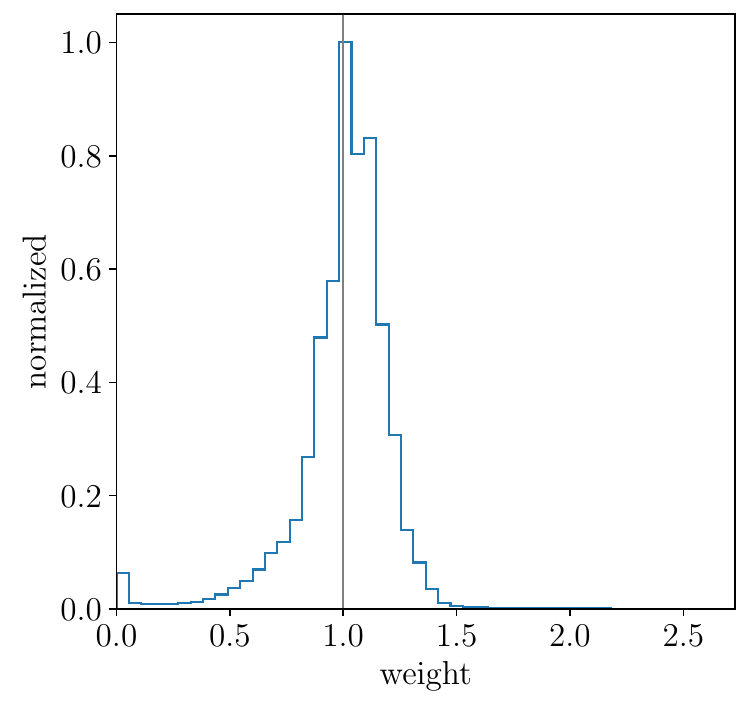}
    \caption{Profiled and marginalized likelihoods, 68\% and 95\%
      confidence regions, and weight distributions for the Gaussian
      mixture model with $d=4$ (upper) and $d=10$ (lower).}
    \label{fig:fivepoints}
\end{figure}
%----------------------------------------------------------

As a second toy model consider a 2-dimensional Gaussian mixture with a
large central peak and four smaller peaks in the corners,
\begin{align}
    p(x) \propto 10 \times \mathcal{N}(x; \mu=(0,0), \sigma=1) \;
    + \sum_{\mu_{1,2}=\pm d} \mathcal{N}(x; \mu=(\mu_1,\mu_2), \sigma=0.5) \; .
\end{align}
We start with the more compact spacing $d = 4$ and show the results in the upper
panels of Fig.~\ref{fig:fivepoints}. Even when we set the distribution
of the initial samples during pre-scaling to the width of the central
peak, we find that the ML-sampling maps out all four outer peaks.

This gets more difficult when we increase the distance from the center
to the outer peaks. Around $d=8$, we start to observe mode collapse,
so some of the outer peaks are no longer found. However, this can be
easily prevented by increasing the spread of the initial samples
during pre-scaling. With that adjustment, the ML-sampling easily learn
the distribution for $d=10$ without either very large or very small
sample weights, as demonstrated in the lower panels of
Fig.~\ref{fig:fivepoints}. We even find that the ML-sampling still
learns the distribution if we make all peaks narrower by a factor of
ten, albeit at the cost of more samples with very small weights.

%\clearpage
%%%%%%%%%%%%%%%%%%%%%%%%%%%%%%%%%%%%%%%%%%%%%%%%%%%
\subsection{\sfitter likelihood}
\label{sec:res_lhc}

Moving on to the LHC, we begin by studying the individual sectors separately to prove the validity of our method, while simultaneously presenting the improvements we gain from having access to the gradients of our likelihood. Finally, we display the full strength of our methods by applying them to the much higher dimensional combined analysis. In particular, we focus on the improvements on the required computational time compared to previous \sfitter analyses.

%%%%%%%%%%%%%%%%%%%%%%%%%%%%%%%%%%%%%%%%%%%%%%%%%%%
\subsubsection*{Top sector}

%----------------------------------------------------------
\begin{figure}[b!]
    \includegraphics[width = 0.495\textwidth]{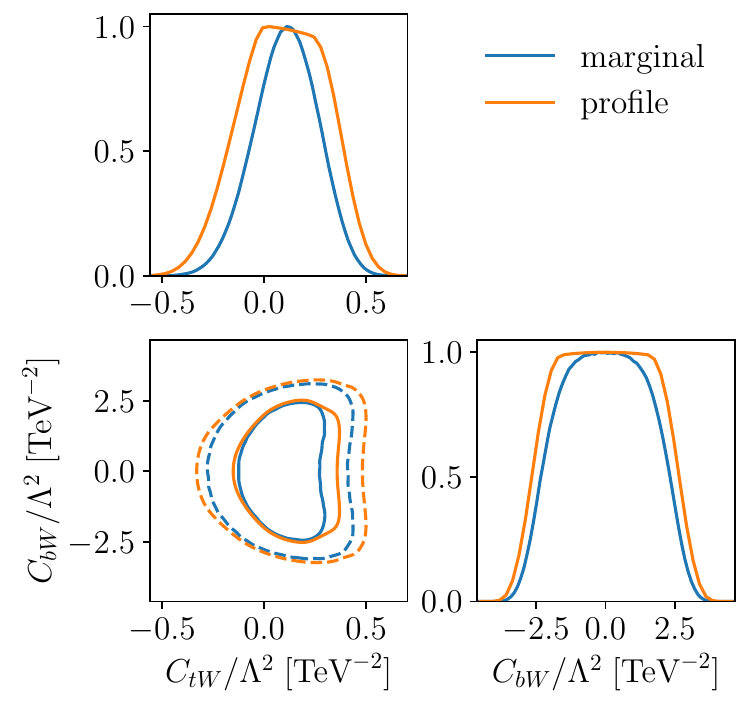}
    \includegraphics[width = 0.495\textwidth]{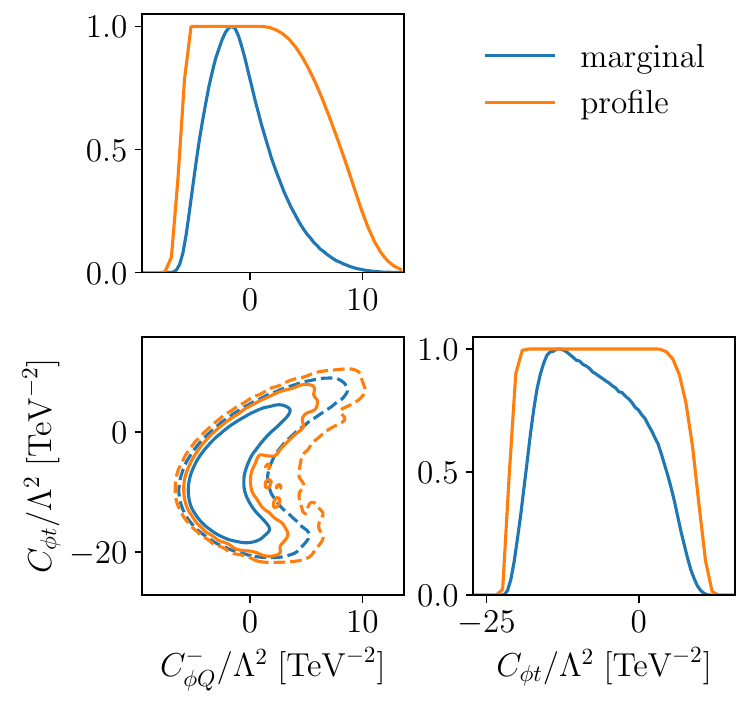}\\
    \includegraphics[width = 0.495\textwidth]{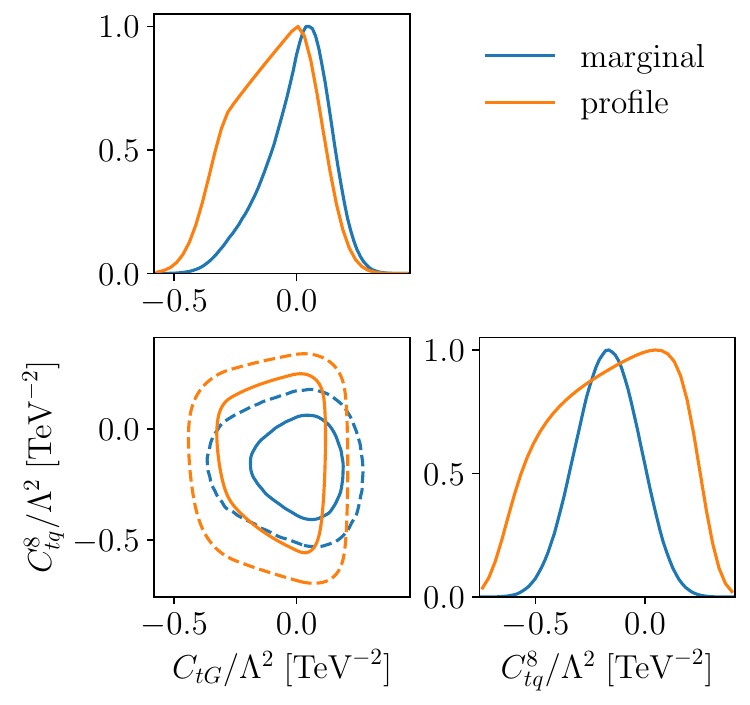}
    \includegraphics[width = 0.495\textwidth]{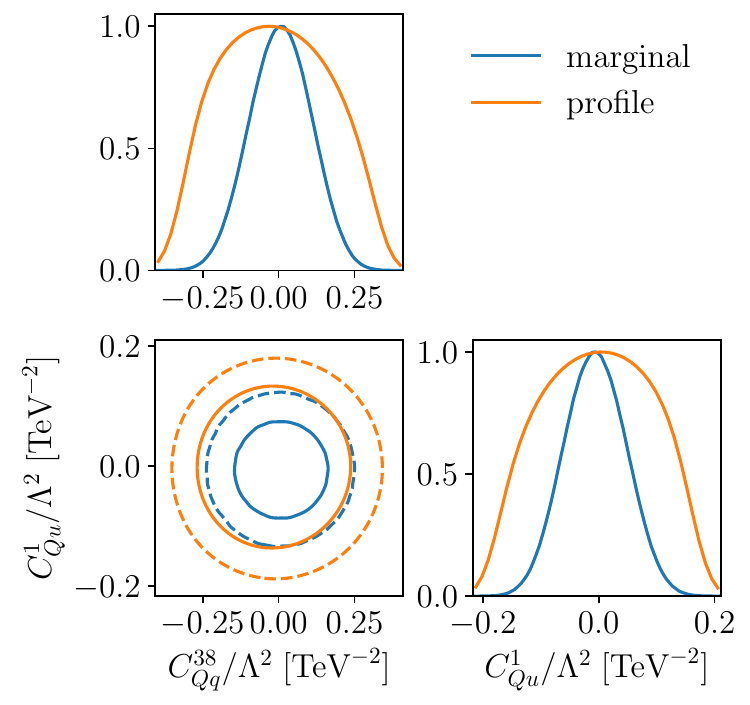}
    \caption{Correlations between select Wilson coefficients in the Top sector, showing results after both profiling (orange) and marginalizing (blue) over the remaining Wilson coefficients.}
    \label{fig:top_2d}
\end{figure}
%----------------------------------------------------------

We begin with the top sector, since its likelihood should be particularly simple and as such provides the perfect environment to study some basic features and technical improvements. For reference, all correlations for the 22 Wilson coefficients can be found in Fig.~\ref{fig:happy_top_2d} in the Appendix. 

In Fig.~\ref{fig:top_2d} we show correlations for a few selects pairs of Wilson coefficients which display interesting patterns. In the upper left panel we show the correlations between $C_{tW}$ and $C_{bW}$, with a slight correlation and the typical flat profile likelihoods induced by theory uncertainties. While the 1-dimensional profiled and marginalized likelihoods have a similar width, the marginalization leads to a well-defined maximum and round edges by construction. The upper right panels illustrates a non-trivial correlation between $C_{\phi Q}^{-}$ and $C_{\phi t}$, perfectly reproducing earlier \sfitter results. Because of the form of the correlation, the 1-dimensional limits on $C_{\phi Q}^{-}$ after marginalization are much stronger than after profiling. This reflects a sizeable impact of the implicit bias by integrating over the Wilson coefficients linearly.  The bottom left panel shows the correlations between $C_{tG}$ and $C_{tq}^{8}$ where we see a secondary, but non-degenerate likelihood structure leading to the enhanced shoulder in both 1-dimensional profile likelihood. The marginalization washes out this effect. Finally, we show the correlation between the two four-fermion operators $C_{Qq}^{38}$ and $C_{Qu}^{1}$. Here, we clearly see the effect of the large theory uncertainties on the top rate measurements, providing much tighter constraints after marginalizing, even though the flat behavior of the profile likelihood is washed out by the additional, correlated directions already profiled away.

%----------------------------------------------------------
\begin{table}[t]
    \centering
    \begin{small} \begin{tabular}[t]{llll}
    \toprule
    & Top & Higgs-gauge & Combined \\
    \midrule
    Dimensions       & 22 & 20 & 42 \\
    Training batches & 100 & 2000 & 6000 \\
    Samples          & 10M & 200M & 100M \\
    \midrule
    Effective sample size & 7.1M & 97M & 21M \\
    \midrule
    Pre-scaling time  & 7s & 3.5min & 5.3min \\
    Pre-training time & 18s & 1.7min & 2.5min \\
    Training time     & 36s & 17.3min & 1.2h \\
    Sampling time     & 26s & 14.8min & 17.6min \\
    Profiling time    & 17.7min & 24.8min & 3.7h \\
    \midrule
    Number of CPUs & 20 & 80 & 120 \\
    Accepted samples & 37M & 26.4M & 60M \\
    CPU sampling time & 29min 49s & 3h 23min & 20h 50min \\
    CPU profiling time & 4min 43s & 8min 24s & N/A \\
    \bottomrule
    \end{tabular} \end{small}
    \caption{Training and sampling time for the five steps to happiness on an H100 GPU. The CPU times refer to the time needed on the given number of CPUs for the original analyses~\cite{Brivio:2022hrb,Elmer:2023wtr} on the NEMO Cluster. (CPU profiling time for a single 2D correlation plot on a single CPU.)}
    \label{tab:performance}
\end{table}
%----------------------------------------------------------

The time required to perform the full analysis can be found in the left column of Tab.~\ref{tab:performance}, where also the time needed to perform a traditional \sfitter top analysis is listed. Due to the simplicity of the likelihood, both methods generate many samples efficiently. The entire analysis, including the profiling, runs in minutes on the GPU, and the main difference to CPU-based \sfitter studies is that now the profiling and marginalization lead to equally smooth distributions. Within \sfitter we never achieved this equivalent smoothness, even using 20 CPUs in parallel.
%this requires additional profiling time, as listed in Tab.~\ref{tab:performance}, while providing less smooth results. 

%%%%%%%%%%%%%%%%%%%%%%%%%%%%%%%%%%%%%%%%%%%%%%%%%%%
\subsubsection*{Higgs-gauge sector}

\begin{figure}[t]
    \centering
    \includegraphics[width = 0.48\textwidth]{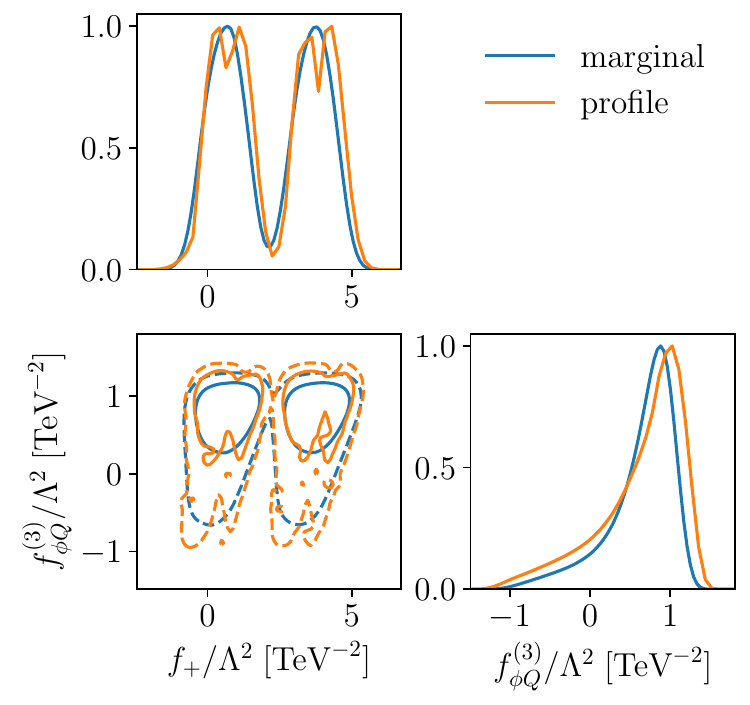}
    \includegraphics[width = 0.48\textwidth]{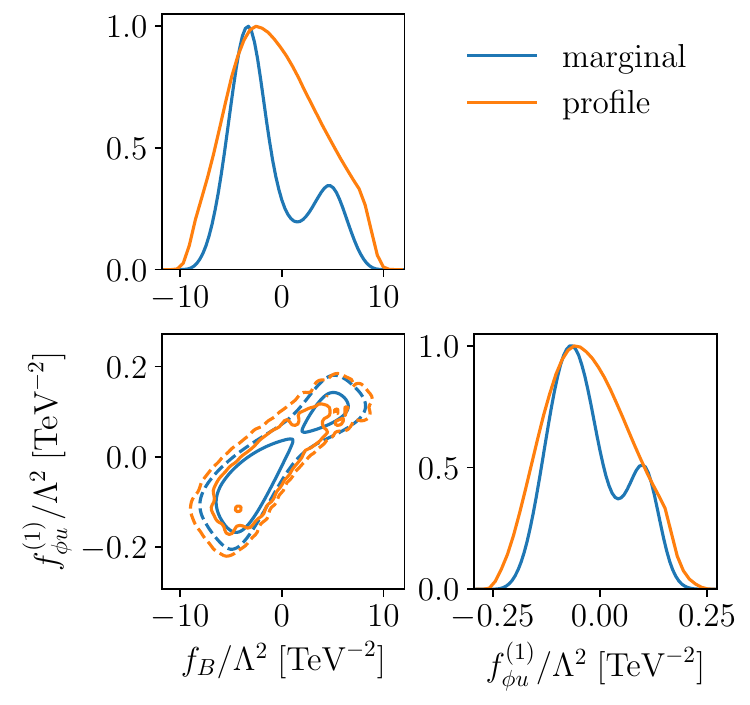}
    \caption{Correlations between two sets of Wilson coefficients in the Higgs sector after profiling (orange) and marginalization (blue).}
    \label{fig:higgs_2d}
\end{figure}

In the Higgs-gauge sector, the likelihood evaluation becomes harder, because of the more complex and highly correlated shape of the likelihood. This leads to numerical challenges, even though the dimensionality is similar to that of the top sector. Fig.~\ref{fig:higgs_2d} shows two example combinations of Wilson coefficients, while all other correlations can again be found within in Fig.~\ref{fig:happy_higgs_2d} of the Appendix. At this point we use the operator set in the HISZ basis to allow for an easier comparison with previous \sfitter results.

Both plots illustrate the more complicated shape of our likelihood, with two modes now appearing for various Wilson coefficients.
In the left panel we see the correlations between $f_{+}$ and $f_{\phi Q}^{(3)}$ where we see two distinct modes for $f_{+}$ after profiling and after marginalization. Clearly, one mode is incompatible with small deviations from the SM, and in previous \sfitter analyses the final global analysis was restricted to the SM-like mode. Similarly, all secondary modes arising from sign flips in the Yukawa couplings are removed in the final constraints, since these would require large new physics effects one would likely have observed already.

In the right panels we show the correlation between $f_{B}$ and $f_{\phi u}^{(1)}$ where after marginalization we again find multiple modes for the 1D distributions of our Wilson coefficients while the profiled results do not show this structure. We observe a dip at the SM for both of these coefficients, the cause of which was examined and discussed in great detail in Ref.~\cite{Brivio:2022hrb}. To clarify briefly, these differences are the result of an underfluctuation in a kinematic distribution which needs to be explained in the SMEFT. During profiling we always select the most likely point in parameter space, this is typically close to the SM and independent of the dimensionality of the WC parameter space. When marginalizing, however, volume effects appear from the larger space of Wilson coefficients, creating the two peaks at non-SM values.

%----------------------------------------------------------
%----------------------------------------------------------

The timing numbers in the central column of Tab.~\ref{tab:performance} reflect the numerical challenges from the more complicated correlated Higgs-gauge likelihood. The effective sample size is much larger than for the top sector, and the sampling time for the GPU implementation is also much longer. A full \sfitter analysis needs a total of 80 CPUs for a few hours, and the shape of the profile likelihood of the published results were often quite rugged. Comparatively, we find that even for the more complicated Higgs likelihood our new results are much smoother.

%%%%%%%%%%%%%%%%%%%%%%%%%%%%%%%%%%%%%%%%%%%%%%%%%%%
\subsection*{Combined analysis}

\begin{figure}[t]
    \centering
    \includegraphics[width = \textwidth]{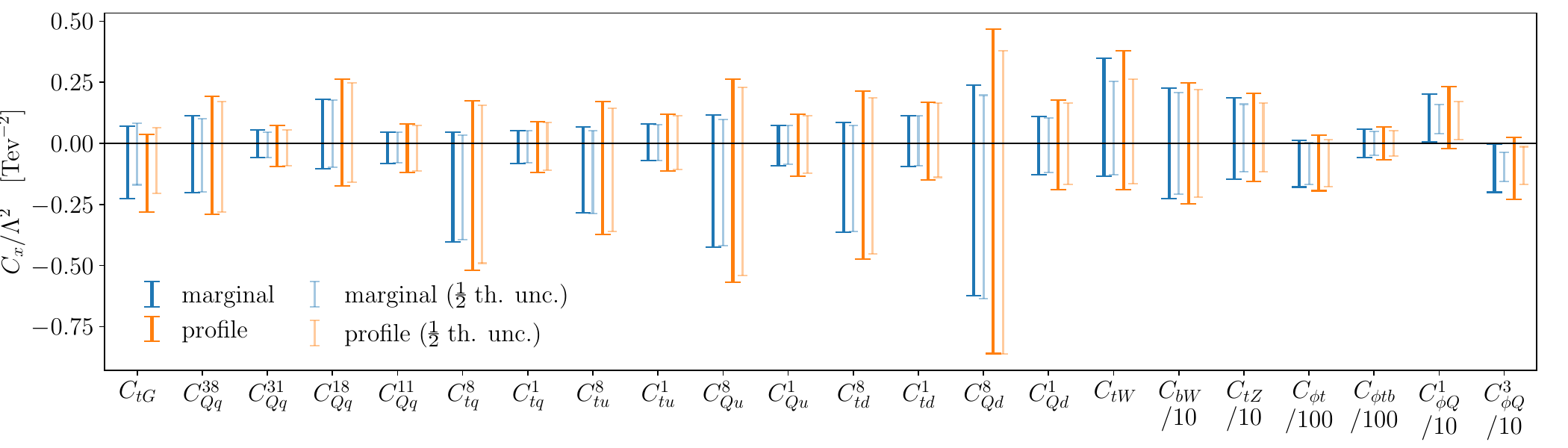}\\
    \includegraphics[width = \textwidth]{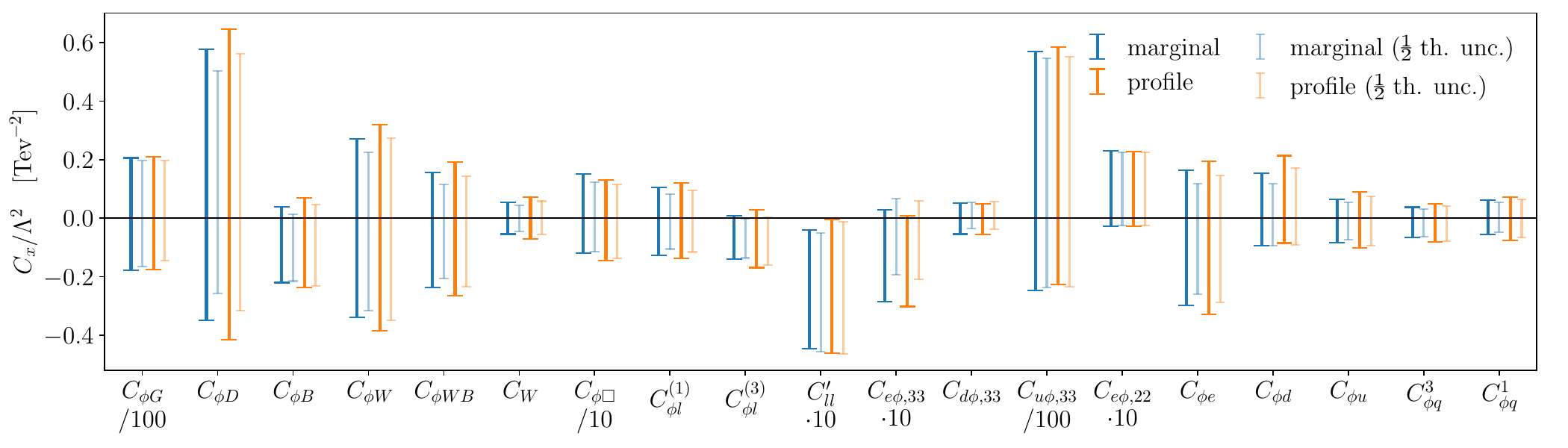}
    \caption{Constraints from the combined SMEFT analysis in the top (top) and Higgs-gauge (bottom) sectors, showing 95\% CLs for all 42 Wilson coefficients for both profiled and marginalized likelihoods.}
    \label{fig:money}
\end{figure}

The final step in our analysis is to combine both datasets and perform a full global analysis of all 42 Wilson coefficients. The detailed results in the Warsaw basis can be found in Fig.~\ref{fig:happy_combined_2d}. In Fig.~\ref{fig:money} we summarize the results of all these coefficients, split up between the coefficients of the Top sector at the top and those for the Higgs-gauge sector at the bottom. The 1-dimensional profiled and marginalized likelihoods are given in Fig.~\ref{fig:combined_1d} of the Appendix. In addition to the current limits from Run 2, we also show hypothetical limits assuming that we could reduce the theory uncertainty by a factor of two, leaving the central predictions the same. Because tails of kinematic distributions are statistics-limited, and because there is no tension in the global analysis, the impact of such an improvement is limited to a few Wilson coefficients, like $C_{\phi Q}^1$ or $C_{\phi B}$. 

As before, we look at the training and sampling times listed in Tab.~\ref{tab:performance}. While the increased dimensionality of the likelihood requires serious training, the sampling is faster by many orders of magnitude compared to our previous \sfitter studies on 120 CPUs in parallel. The relative speed improvement for sampling on a single GPU is a factor 68 for 120 CPUs vs a single GPU and 8200 for a single CPU vs a single GPU. A complete set of profile likelihoods were, essentially, out of numerical reach for the current CPU implementation.

%----------------------------------------------------------
%----------------------------------------------------------

%%%%%%%%%%%%%%%%%%%%%%%%%%%%%%%%%%%%%%%%%%%%%%%%%%%
%\subsection*{TODO list}
%\begin{itemize}
%\item top sector: D6tw vs D6bw for theory, D6phit vs D6phim for theory, D6tg vs D6qt8 for almost-generate minima, for theory marginalized, D6qu1 vs D6qq83 for just blowing up.
%\item Higgs-gauge sector: top vs gluon for correlation, minus vs fb for slightly broken degeneracy (maybe Warsaw basis?)
%\item combined: tg vs something else, to show that nothing has changed...
%\item combined: 1D of everything, money plot
%\end{itemize}

%%%%%%%%%%%%%%%%%%%%%%%%%%%%%%%%%%%%%%%%%%%%%%%%%%%
\section{Outlook}
\label{sec:outlook}

Global SMEFT analyses are a powerful method telling us to what level the LHC results as a whole agree with the Standard Model (or not). They cover rate measurements and kinematic distributions, defining a large number of measurements with many experimental and theory uncertainties, some of them correlated. Once we combine, for instance, the Higgs-gauge sector with the top sector, the number of Wilson coefficients becomes sizeable as well. This means that the numerical extraction and the analysis of the fully exclusive likelihood is a numerical challenge. 

We have shown that applying ML-techniques to learn and evaluate the likelihood allows us to run a global \sfitter analysis on a single GPU in a few hours, rather than on a CPU cluster for a day or more. Our five steps to happiness are inspired by neural importance sampling and MadNIS~\cite{Heimel:2022wyj,Heimel:2023ngj,Heimel:2024wph}. They include (i) pre-scaling of the likelihood parameters; (ii) pre-training using a normalizing flow, (iii) training using annealed importance sampling and buffered training, (iv) sampling from the normalizing likelihood flow; and (v) using gradients for efficient profiling. Especially the last step does not only speed up the likelihood evaluation, it also increases its numerical resolution significantly. For future \sfitter analyses, built on a comprehensive and correlated uncertainty treatment, these numerical improvements mean transformative progress.

%%%%%%%%%%%%%%%%%%%%%%%%%%%%%%%%%%%%%%%%%%%%%%%%%%%
\subsection*{Acknowledgements}

This research is supported through the KISS consortium (05D2022)
funded by the German Federal Ministry of Education and Research BMBF
in the ErUM-Data action plan, by the Deutsche Forschungsgemeinschaft
(DFG, German Research Foundation) under grant 396021762 -- TRR~257:
\textsl{Particle Physics Phenomenology after the Higgs Discovery}, and
through Germany's Excellence Strategy EXC~2181/1 -- 390900948 (the
\textsl{Heidelberg STRUCTURES Excellence Cluster}).  TH was funded by
the Carl-Zeiss-Stiftung through the project \textsl{Model-Based AI:
  Physical Models and Deep Learning for Imaging and Cancer Treatment}.
The authors acknowledge support by the state of Baden-Württemberg
through bwHPC and the German Research Foundation (DFG) through grant
no INST 39/963-1 FUGG (bwForCluster NEMO).

%\clearpage
%%%%%%%%%%%%%%%%%%%%%%%%%%%%%%%%%%%%%%%%%%%%%%%%%%%
\appendix
\section{2-dimensional correlations}

%----------------------------------------------------------
\begin{figure}[b!]
    \centering
    \includegraphics[width = \textwidth, page = 1]{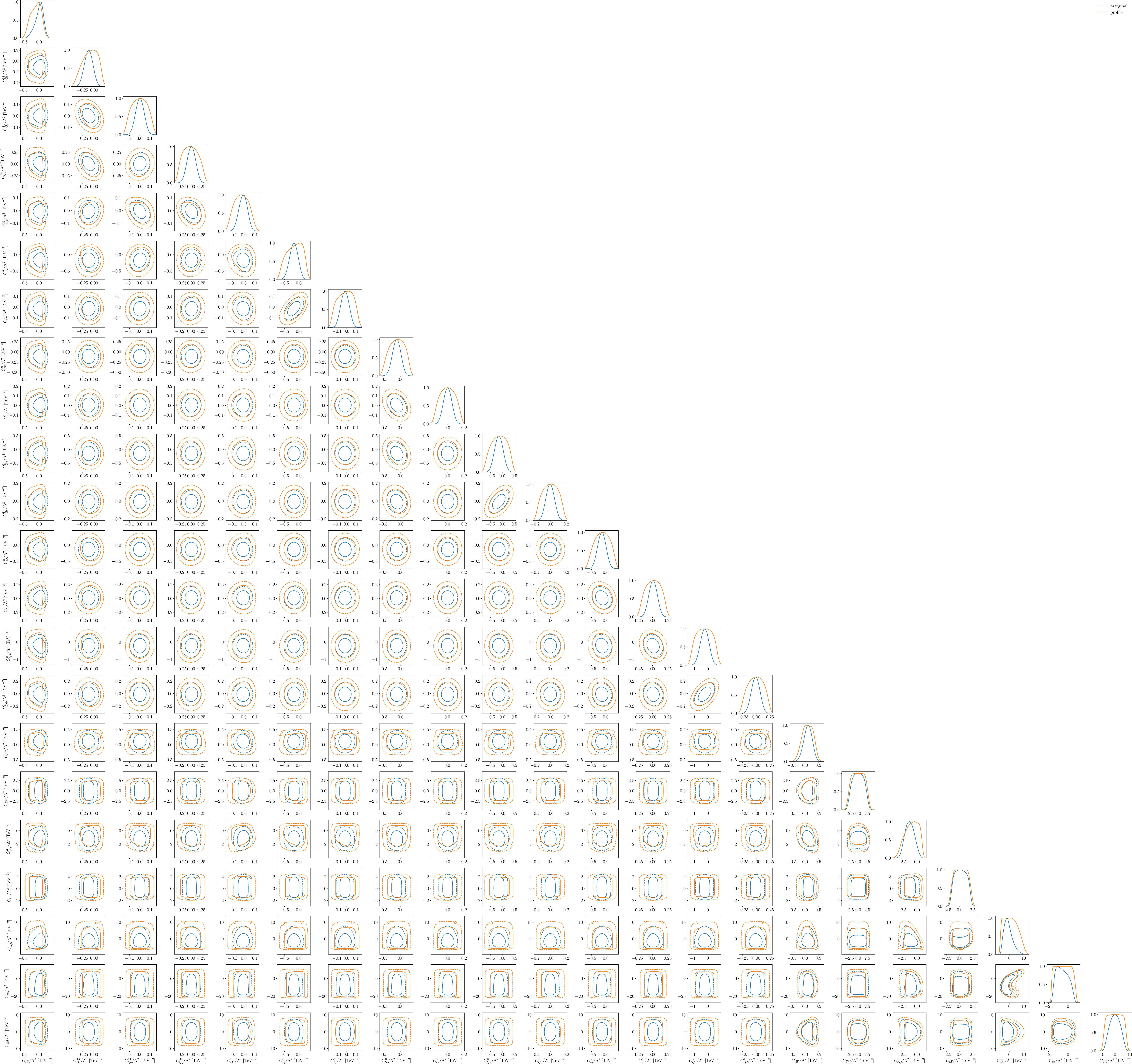}
    \caption{2-dimensional and 1-dimensional profile and marginalized likelihoods for the top sector.}
    \label{fig:happy_top_2d}
\end{figure}
%----------------------------------------------------------

%----------------------------------------------------------
\begin{figure}
    \centering
    \includegraphics[width = \textwidth, page = 1]{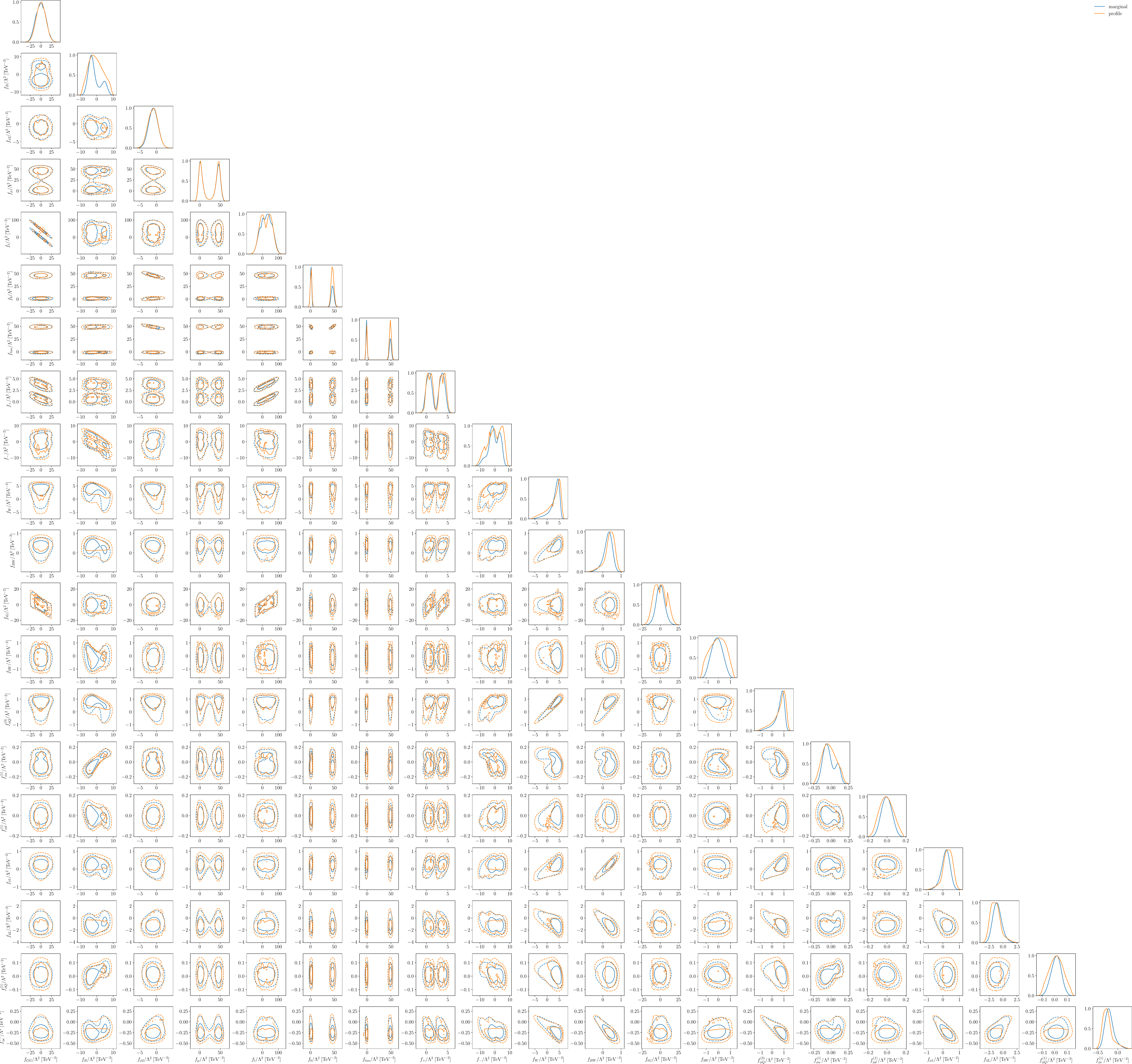}
    \caption{2-dimensional and 1-dimensional profile and marginalized likelihoods for the Higgs-gauge sector in the HISZ operator basis.}
    \label{fig:happy_higgs_2d}
\end{figure}
%----------------------------------------------------------

%----------------------------------------------------------
\begin{figure}
    \centering
    \includegraphics[width = \textwidth, page = 1]{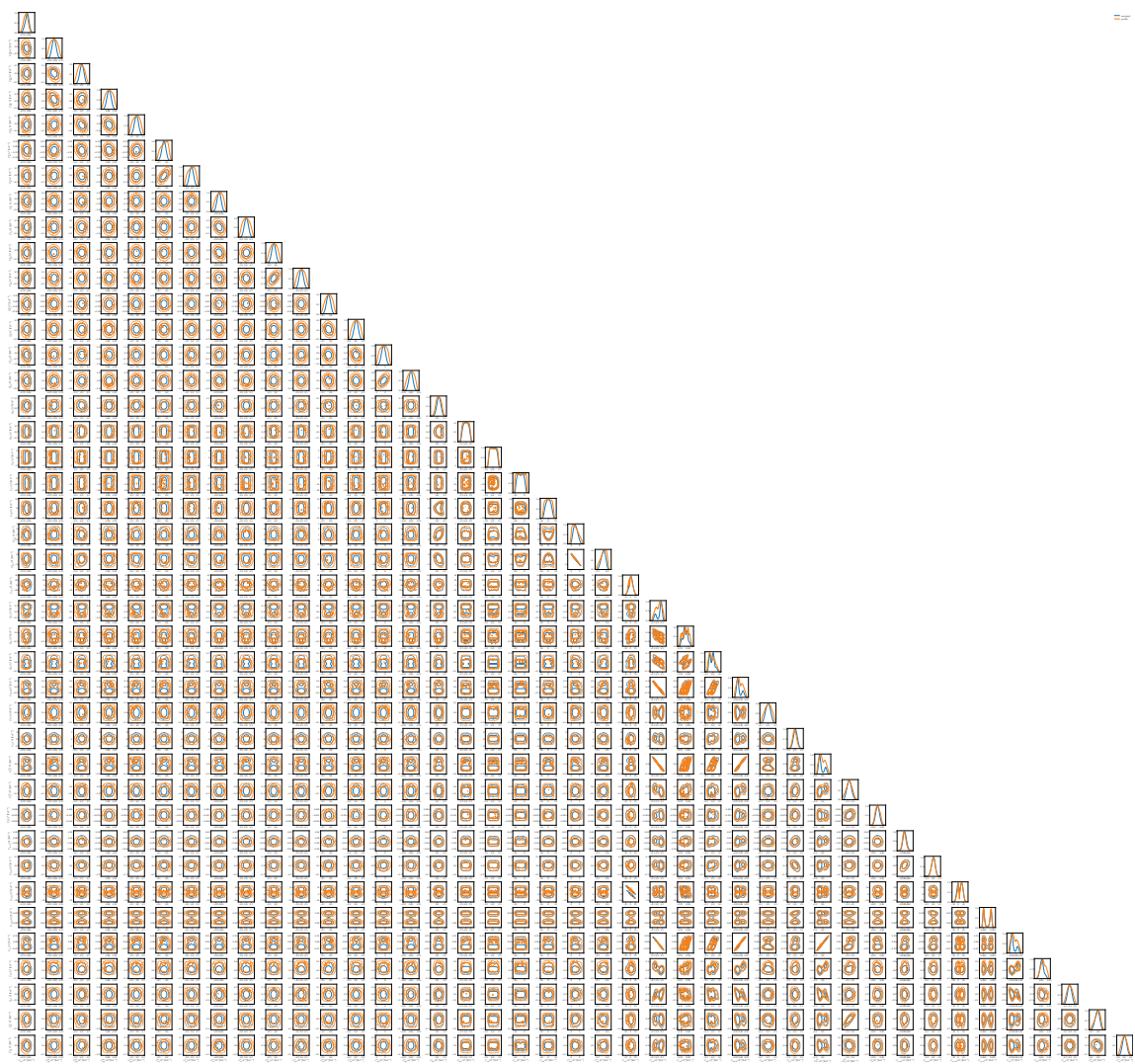}
    \caption{2-dimensional and 1-dimensional profile and marginalized likelihoods for the Higgs-gauge and top sectors combined.}
    \label{fig:happy_combined_2d}
\end{figure}
%----------------------------------------------------------

%----------------------------------------------------------
\begin{figure}
    \includegraphics[width = 0.16\textwidth, page = 1]{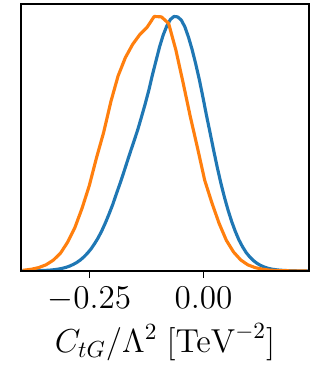}
    \includegraphics[width = 0.16\textwidth, page = 2]{figs/combined/combined_hist1d.pdf}
    \includegraphics[width = 0.16\textwidth, page = 3]{figs/combined/combined_hist1d.pdf}
    \includegraphics[width = 0.16\textwidth, page = 4]{figs/combined/combined_hist1d.pdf}
    \includegraphics[width = 0.16\textwidth, page = 5]{figs/combined/combined_hist1d.pdf}
    \includegraphics[width = 0.16\textwidth, page = 6]{figs/combined/combined_hist1d.pdf}\\
    \includegraphics[width = 0.16\textwidth, page = 7]{figs/combined/combined_hist1d.pdf}
    \includegraphics[width = 0.16\textwidth, page = 8]{figs/combined/combined_hist1d.pdf}
    \includegraphics[width = 0.16\textwidth, page = 9]{figs/combined/combined_hist1d.pdf}
    \includegraphics[width = 0.16\textwidth, page = 10]{figs/combined/combined_hist1d.pdf}
    \includegraphics[width = 0.16\textwidth, page = 11]{figs/combined/combined_hist1d.pdf}
    \includegraphics[width = 0.16\textwidth, page = 12]{figs/combined/combined_hist1d.pdf}\\
    \includegraphics[width = 0.16\textwidth, page = 13]{figs/combined/combined_hist1d.pdf}
    \includegraphics[width = 0.16\textwidth, page = 14]{figs/combined/combined_hist1d.pdf}
    \includegraphics[width = 0.16\textwidth, page = 15]{figs/combined/combined_hist1d.pdf}
    \includegraphics[width = 0.16\textwidth, page = 16]{figs/combined/combined_hist1d.pdf}
    \includegraphics[width = 0.16\textwidth, page = 17]{figs/combined/combined_hist1d.pdf}
    \includegraphics[width = 0.16\textwidth, page = 18]{figs/combined/combined_hist1d.pdf}\\
    \includegraphics[width = 0.16\textwidth, page = 19]{figs/combined/combined_hist1d.pdf}
    \includegraphics[width = 0.16\textwidth, page = 20]{figs/combined/combined_hist1d.pdf}
    \includegraphics[width = 0.16\textwidth, page = 21]{figs/combined/combined_hist1d.pdf}
    \includegraphics[width = 0.16\textwidth, page = 22]{figs/combined/combined_hist1d.pdf}
    \includegraphics[width = 0.16\textwidth, page = 23]{figs/combined/combined_hist1d.pdf}
    \includegraphics[width = 0.16\textwidth, page = 24]{figs/combined/combined_hist1d.pdf}\\
    \includegraphics[width = 0.16\textwidth, page = 25]{figs/combined/combined_hist1d.pdf}
    \includegraphics[width = 0.16\textwidth, page = 26]{figs/combined/combined_hist1d.pdf}
    \includegraphics[width = 0.16\textwidth, page = 27]{figs/combined/combined_hist1d.pdf}
    \includegraphics[width = 0.16\textwidth, page = 28]{figs/combined/combined_hist1d.pdf}
    \includegraphics[width = 0.16\textwidth, page = 29]{figs/combined/combined_hist1d.pdf}
    \includegraphics[width = 0.16\textwidth, page = 30]{figs/combined/combined_hist1d.pdf}\\
    \includegraphics[width = 0.16\textwidth, page = 31]{figs/combined/combined_hist1d.pdf}
    \includegraphics[width = 0.16\textwidth, page = 32]{figs/combined/combined_hist1d.pdf}
    \includegraphics[width = 0.16\textwidth, page = 33]{figs/combined/combined_hist1d.pdf}
    \includegraphics[width = 0.16\textwidth, page = 34]{figs/combined/combined_hist1d.pdf}
    \includegraphics[width = 0.16\textwidth, page = 35]{figs/combined/combined_hist1d.pdf}
    \includegraphics[width = 0.16\textwidth, page = 36]{figs/combined/combined_hist1d.pdf}\\
    \includegraphics[width = 0.16\textwidth, page = 37]{figs/combined/combined_hist1d.pdf}
    \includegraphics[width = 0.16\textwidth, page = 38]{figs/combined/combined_hist1d.pdf}
    \includegraphics[width = 0.16\textwidth, page = 39]{figs/combined/combined_hist1d.pdf}
    \includegraphics[width = 0.16\textwidth, page = 40]{figs/combined/combined_hist1d.pdf}
    \includegraphics[width = 0.16\textwidth, page = 41]{figs/combined/combined_hist1d.pdf}
    \includegraphics[width = 0.16\textwidth, page = 42]{figs/combined/combined_hist1d.pdf}
    \caption{1D profiled (orange) and marginalized (blue) likelihoods for the combined fit.}
    \label{fig:combined_1d}
\end{figure}
%----------------------------------------------------------

\clearpage
%%%%%%%%%%%%%%%%%%%%%%%%%%%%%%%%%%%%%%%%%%%%%%%%%%%
\section{Hyperparameters}

%----------------------------------------------------------
\begin{table}[h!]
    \centering
    \begin{small} \begin{tabular}[t]{lllll}
    \toprule
    && Top & Higgs-gauge & Combined \\
    \midrule
    Architecture
    & Coupling blocks & RQ splines && \\
    & Spline bins     & 16 && \\
    & Subnet layers   & 3 && \\
    & Hidden layers   & 64 && \\
    \midrule
    Pre-scaling
    & Number of samples & 10240 & 40960 & 40960 \\
    & AIS steps         & 1500 & 5500 & 5500 \\
    & Target acceptance & 0.33 && \\
    \midrule
    Pre-training
    & Batch size                 & 1024 && \\
    & Epochs                     & 15 & 6 & 6 \\
    & MCMC steps between batches & 20 & 10 & 10 \\
    \midrule
    Training
    & Learning rate               & 0.001 && \\
    & Batch size                  & 1024 && \\
    & Batches                     & 100 & 2000 & 6000 \\
    & AIS steps                   & 4 & 4 & 8 \\
    & Buffer capacity             & 262k && \\
    & Ratio buffered/online steps & 6 && \\
    \midrule
    Sampling
    & Batches & 100 & 2000 & 1000 \\
    & Batch size & 100k && \\
    & Marginalization bins, 1D & 80 && \\
    & Marginalization bins, 2D & 40 && \\
    & Profiling bins, 1D & 40 && \\
    & Profiling bins, 2D & 30 & 30 & 20 \\
    \midrule
    Profiling
    & Batch size         & 100k && \\
    & Optimizer          & LBFGS && \\
    & Optimization steps & 200 && \\
    \bottomrule
    \end{tabular} \end{small}
    \caption{Hyperparameters for the five fitting steps. If only one value is given, it applies to all three fits.}
    \label{tab:hyperparameters}
\end{table}
%----------------------------------------------------------

%%%%%%%%%%%%%%%%%%%%%%%%%%%%%%%%%%%%%%%%%%%%%%%%%%
\bibliography{tilman,refs,literature,top-eft,toplikelihoods}
\end{document}